\documentclass[lettersize,journal]{IEEEtran}
\usepackage{amsmath,amsfonts}
\usepackage{array}
\usepackage{textcomp}
\usepackage{stfloats}
\usepackage{braket}
\usepackage{url}
\usepackage{verbatim}
\usepackage{graphicx}
\usepackage{cite}
\usepackage{mathtools}
\usepackage{amsmath,amsfonts}
\usepackage{array}
\usepackage{multirow}
\usepackage{array, caption}
\usepackage{graphicx}
\usepackage{makecell}
\usepackage{amssymb, nccmath}
\usepackage{textcomp}
\usepackage{url}
\usepackage{subfigure}
\usepackage{tikz}
\usepackage{pgfplots}
\usepackage{tikz}
\usepackage{ragged2e}
\usepackage{amsmath,amsfonts}
\usetikzlibrary{patterns}
\definecolor{myhund}{HTML}{BE0032}
\definecolor{myfifty}{HTML}{FF003F}
\definecolor{mycolor1}{HTML}{F5F5DC}
\definecolor{mytwenty}{HTML}{8B008B}
\definecolor{myzero}{HTML}{FF007F}
\definecolor{pssfhun}{HTML}{0000FF}
\definecolor{pssffif}{HTML}{333399}
\definecolor{pssftwen}{HTML}{8A2BE2}
\definecolor{pssfzero}{HTML}{6699CC}
\definecolor{ff}{HTML}{D2691E}
\definecolor{bf}{HTML}{FFBF00}
\definecolor{rf}{HTML}{967117}
\definecolor{s1}{HTML}{CE2029}
\definecolor{s2}{HTML}{006D5B}
\definecolor{s3}{HTML}{8DB600}
\definecolor{r4}{HTML}{CE2029}
\usepackage{graphicx}

\usepackage{amsthm}
\usepackage{amssymb}
\usepackage[linesnumbered, ruled]{algorithm2e}
\makeatletter
\newcommand{\removelatexerror}{\let\@latex@error\@gobble}
\makeatother
\usepackage{dsfont}
\usepackage{cite}
\makeatletter
\def\ps@IEEEtitlepagestyle{%
	\def\@oddfoot{\mycopyrightnotice}%
	\def\@oddhead{\hbox{}\@IEEEheaderstyle\leftmark\hfil\thepage}\relax
	\def\@evenhead{\@IEEEheaderstyle\thepage\hfil\leftmark\hbox{}}\relax
	\def\@evenfoot{}%
}

\def\mycopyrightnotice{%
	\begin{minipage}{\textwidth}
		\centering \scriptsize
		© 2025 IEEE. This article has been accepted in IEEE Transactions on Systems, Man, and Cybernetics Journal © 2025 IEEE. Personal use of this material is permitted. Permission from IEEE must be obtained for all other uses, in any current or future media, including reprinting/republishing this material for advertising or promotional purposes, creating new collective works, for resale or redistribution to servers or lists, or reuse of any copyrighted component of this work in other works. This work is freely available for survey and citation.		
	\end{minipage}
}
\makeatother
\newtheorem{theorem}{Theorem}
\newtheorem{corollary}{Corollary}{}
\newtheorem{definition}{Definition}[]

\begin{document}
\title{Quantum Blackhole Learning-Optimized Hadamard Neural Network Model for Dynamic Resource Reservation in Industry Clouds }
\author{Deepika~Saxena,~\textit{Member, IEEE},~Hari~Mohan~Gaur,~\textit{Member, IEEE},\\~Ashutosh~Kumar~Singh, ~\textit{Senior Member, IEEE},~Anand~Mohan,~\textit{Senior Member, IEEE}
	\IEEEcompsocitemizethanks{\IEEEcompsocthanksitem D. Saxena is with School of Computer Science and Engineering, The University of Aizu, Japan and also with Department of Computer Science, Vizja University, Warsaw, Poland.  E-mail: deepika@u-aizu.ac.jp,  13deepikasaxena@gmail.com\\
	H. M. Gaur is with	Institute of Computer Science, Goethe University, Frankfurt, Germany. E-mail: leoharimohan84@gmail.com 
		 \\	
		 A. K. Singh is  with Indian Institute of Information Technology Bhopal, India and also with Department of Computer Science, Vizja University, Warsaw, Poland, Europe. E-mail:  ashutosh@iiitbhopal.ac.in\\
	A. Mohan is with Department of Electronics Engineering
	Indian Institute of Technology (BHU), India. E-mail: profanandmohan@gmail.com
		
}}




\maketitle

\begin{abstract}
Accurate workload prediction and proactive resource reservation are crucial for industry clouds. However, the conventional machine learning models with limited learning capabilities often fail to predict diverse, high-dimensional workloads with sudden changes in resource demand, leading to excessive power consumption and resource management issues. In this context, this paper proposes  a novel Hadamard Neural Network with Quantum Blackhole (QB-HNN) optimization. This model combines the computational efficiency of quantum mechanics with the persuasive learning capability of neural networks.   The workload information is transformed into Qubits and propagated via deep network of Qubit neurons comprising of Hadamard gated activation function to fetch superposition  within the QB-HNN model for intuitive pattern learning. Furthermore, a novel Quantum Blackhole Biphase Optimization (QB-BiO)  algorithm is introduced to train and optimize Qubit neural weights. The performance of the proposed model is comprehensively evaluated and compared with five state-of-the-art approaches using six benchmark datasets of three heterogeneous variety of cloud workloads. The prediction accuracy achieved for extensive range of workloads confirms its influential performance by minimizing the prediction error up to 36.36\% and 22.83\% over existing LSTM and EQNN-based prediction approaches, respectively.

\end{abstract}


\section{Introduction}
\IEEEPARstart{T}{he}  dynamic upsurge and plunge of the cloud workloads and resource usage stave off the quality of service, degrades performance, resulting into outages, under-/over-resource utilization, and excessive energy consumption  \cite{tuli2021start,saxena2023performance,11045156}. The foresight reservation of resources by analysing the approximate demand of cloud workloads provides an effective solution to conquer the aforesaid challenges  and establish rational resource provisioning by assorting and allocating the resources adaptively \cite{saxena2022high, saxena2020auto}. 
Therefore, an accurate prediction of heterogeneous cloud workloads is a prerequisite which plays a significant role in the management of entire industry cloud infrastructure. 

\subsection{Related work} \label{rw}
The conventional neural network based prediction models have been widely used for accurate prediction of cloud workloads.  Bi et al. \cite{bi2024arima} proposed workload prediction mechanism combining Savitzky–Golay (SG) filter-based smoothing, wavelet decomposition, and ARIMA for precise statistical trend and component forecasting. A decentralized multiagent-based VM allocation and consolidation approach is proposed in \cite{wang2016multiagent} using auction mechanisms and local negotiation to minimize energy consumption while accounting for migration costs. A dynamic cloud workload prediction model using artificial neural network optimized with self-adaptive differential evolution (SaDE) algorithm has been developed in \cite{kumar2018workload}. This model established improved accuracy over a Backpropagation trained neural network by reason of the adoption of evolutionary learning with  multi-dimensional exploration and convergence for optimization of  neural weights. Kumar et al. \cite{kumar2020biphase} have proposed a Biphase-adaptive
Differential Evolution (BaDE) learning based neural network model to forecast resource usage of heterogeneous cloud workloads. Usage of BaDE for neural weights optimization improved the prediction accuracy over SaDE by establishing a balance between exploration and exploitation.

\par Li et al. \cite{li2018qos} proposed an extended Gale–Shapley algorithm for efficient service composition in cloud manufacturing, effectively handling multiple tasks with diverse constraints and outperforming traditional metaheuristics. A fine-grained workload prediction model using LSTM-RNN was introduced in \cite{kumar2018long}, achieving high accuracy in host load forecasting but suffering from high computation costs due to backpropagation across recurrent layers. To address representational efficiency, Chen et al. \cite{chen2019towards} developed a deep learning-based prediction model (L-PAW) employing a top-sparse autoencoder for workload feature extraction.

Reinforcement learning-based online partitioning was utilized in \cite{li2021reinforcement} to adaptively manage dynamic cloud gaming workloads. Tuli et al.\cite{tuli2021start} introduced an LSTM encoder-based model for automatic prediction and mitigation of straggler tasks, accounting for heterogeneous and volatile cloud environments. Ruan et al.\cite{ruan2022cloud} presented a feature-enhanced deep learning model to detect critical turning points in workload traces, surpassing the limitations of traditional methods.  {Furthermore, Kumar et al.\cite{kumar2020ensemble} proposed an ensemble method combining extreme learning machines and blackhole-inspired optimization to improve prediction accuracy. Building on this, a self-directed workload forecasting model (SDWF) was later introduced in \cite{kumar2021self}, leveraging blackhole-based training and error correction to achieve high-precision predictions across multiple real-world traces.}

\subsection{Research Gaps and Motivation for QB-HNN}

Conventional machine learning (CML) approaches~\cite{kumar2018workload, kumar2020biphase, kumar2018long, kumar2020ensemble, kumar2021self} face severe limitations in addressing the high-dimensional, dynamic, and heterogeneous nature of modern cloud workloads.  {These models are typically optimized for narrowly defined tasks, resulting in poor generalization in unseen scenarios or data distribution shifts. Moreover, their reliance on frequent retraining for diverse workloads introduces substantial computational overhead, hampering real-time adaptability and scalability. Most importantly, traditional CML methods often fail to capture complex nonlinear dependencies and intrinsic quantum-level variability inherent in cloud environments~\cite{wei2021deep, tian2023recent, narayanan2000quantum, kouda2004multilayered, singh2021quantum}.}

 {To address these challenges, our previous works proposed the \textit{Evolutionary Quantum Neural Network (EQNN)}~\cite{singh2021quantum} and the \textit{Multiple Controlled Toffoli-driven Adaptive Quantum Neural Network (MCT-AQNN)}~\cite{gupta2024multiple}. Both models utilized adaptive quantum evolution strategies to optimize qubit weight parameters. EQNN employed Controlled-NOT (C-NOT) gates to manipulate qubit states, while MCT-AQNN enhanced this design by incorporating multiple controlled Toffoli gates for more flexible and adaptive state transformations. These advancements led to notable improvements in learning precision and cloud workload prediction accuracy. However, EQNN and MCT-AQNN still faced critical shortcomings, particularly in handling highly volatile and heterogeneous cloud workloads. Their exclusive reliance on evolutionary algorithms for qubit optimization constrained their ability to explore complex high-dimensional solution landscapes. This often led to premature convergence and stagnation, limiting their applicability in dynamic and unpredictable cloud scenarios. These limitations emphasized the need for a more expressive and intelligent quantum optimization paradigm.}

 {To overcome the bottlenecks of prior models, we propose the \textit{Quantum Blackhole Learning-Optimized Hadamard Neural Network (QB-HNN)}, a novel architecture that represents a substantial advancement in quantum neural modeling for cloud environments. QB-HNN incorporates a \textit{Quantum Blackhole Learning} mechanism, inspired by gravitational dynamics, to intelligently guide candidate solutions toward optimal regions. This results in efficient high-dimensional exploration, faster convergence, and reduced susceptibility to local optima. Furthermore, the integration of the \textit{Hadamard gate} enables quantum superposition, allowing parallel evaluation of multiple solution states. This significantly enhances adaptability to heterogeneous, bursty, and dynamic workload patterns common in industrial cloud systems. QB-HNN also features a \textit{dual-phase learning framework}, enabling efficient interaction between quantum and classical modules for dynamic qubit weight optimization and replay buffer utilization. Notably, it supports two functional modes: one configured for \textbf{single-resource prediction}, and another for \textbf{multi-resource forecasting and reservation}, greatly expanding its applicability in real-world datacenter operations. Beyond mere improvements in speed or complexity, QB-HNN addresses core challenges of generalization, convergence, and adaptability. By fusing quantum computing, blackhole-inspired optimization, and hybrid architectural design, it establishes a new foundation for self-adaptive, precise, and resilient workload management in next-generation quantum-cloud systems.}

\subsection{Our Contributions} 
\begin{figure}[!htbp]
 	\centering
 	\includegraphics[width=0.9\linewidth]{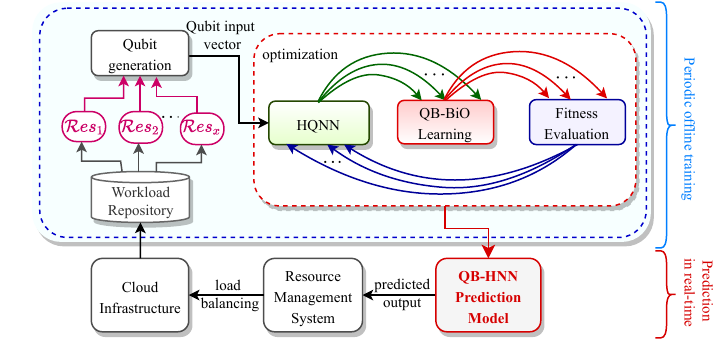}
 	\caption{Schematic view of proposed model }
 	\label{fig:birdev}
 \end{figure}

  A schematic overview of the proposed dynamic resource reservation model is presented in Fig. \ref{fig:birdev} to elucidate the contributions of the proposed model.
It  illustrates a design representation and information flow among intended operational units, wherein the resource usage information \{${R}es_1$, ${R}es_2$, ..., ${R}es_x$\} of the cloud infrastructure stored in the workload repository is pre-processed, and transformed into Qubit input vector (i.e., training data). A novel {H}adamard {Q}uantum {N}eural {N}etwork ({HQNN}) is proposed and optimized by developing and utilizing  Quantum Blackhole Biphase Optimization (QB-BiO) algorithm. During optimization, the fitness is evaluated in terms of prediction error and thus build prediction model is named as \textbf{Q}uantum \textbf{B}lackhole learning-based \textbf{H}adamard \textbf{N}eural \textbf{N}etwork (\textbf{QB-HNN}). The outcome of QB-HNN prediction enables resource management system to reserve the resources in anticipation for the approaching dynamic workloads and manage the resources effectively. The major contributions of the proposed work are as follows:

\begin{itemize}
	\item A novel QB-HNN prediction model is proposed to predict the highly variable demands of  heterogeneous resources and uncertainities of extensive range of cloud workloads. It learns the non-linearities of resource demands and dynamic changes of workloads by leveraging the superposition principle of Hadamard gate and the diversity of qubits with 360 degree rotational capabilities.
	
	\item QB-BiO algorithm is introduced for qubit weight propagation in Quantum Hadamard neural network, which provides a novel approach superior to existing Blackhole optimization. It operates on populations of qubit weights and enhances traditional Blackhole optimization by employing dual optimization capabilities, including \textit{qubit clustering} and \textit{heuristic crossover} techniques.
 
 \item 
Two distinct functional variants of the QB-HNN model are developed to forecast \textit{single-resource demand} (Section \ref{sap}) and \textit{multiple-resource demand} (Section \ref{map}) by adapting the architecture of the QB-HNN model with specific modifications to its learning process.
	
	\item The implementation with a comprehensive evaluation and comparison with the state-of-the-art approaches confirms the potentiality of the proposed model in terms of its accuracy and efficacy  using various workload traces collected from real cloud applications, including cluster, high performance computing (HPC), and web applications.

\end{itemize} 

\textit{Paper Organization}: Section II provides a detailed description of the proposed QB-HNN prediction model, including its comprehensive architecture and two functional variants: \textit{QB-HNN-SAP} (Section \ref{sap}) and \textit{QB-HNN-MAP} (Section \ref{map}). The novel QB-BiO algorithm for qubit weight optimization is explained in Section III. Section IV outlines the experimental setup and presents a performance evaluation, highlighting simulation results and comparative analysis. Finally, conclusive remarks and potential directions for future research are presented in Section V.
\section{QB-HNN Model}
\begin{figure*}[!htbp]
	\centering
	\includegraphics[width=0.76\linewidth]{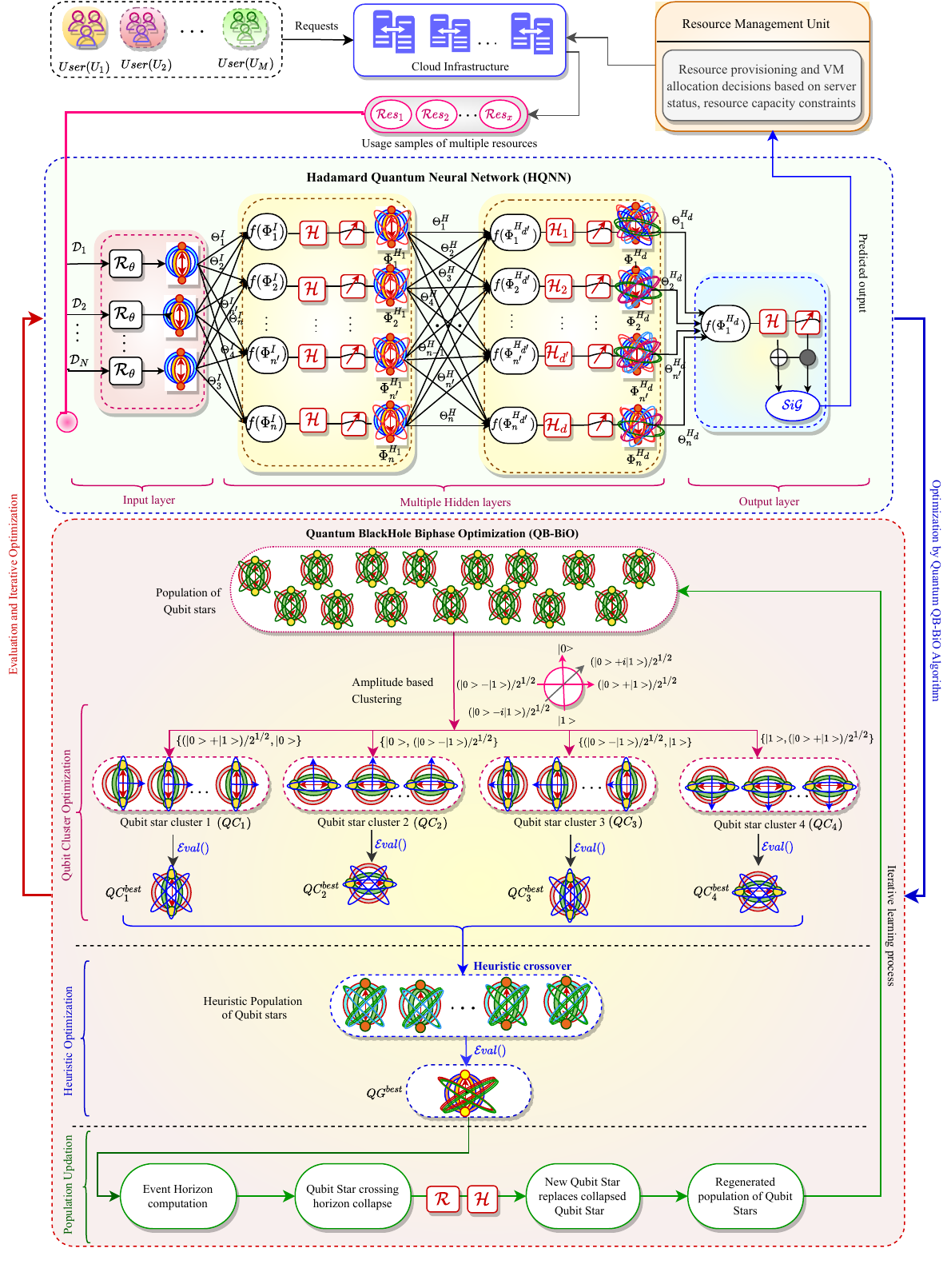}
	\caption{QB-HNN cloud workload prediction model}
	\label{fig:proposedmodel}
\end{figure*}

The architecture of the proposed prediction model is illustrated in Fig. \ref{fig:proposedmodel}.
It depicts essential blocks including \textit{HQNN} and \textit{QB-BiO} based learning of qubit network weights along with the intended information flow among them. Let the $M$ cloud users \{$U_1$, $U_2$, ..., $U_M$\} $\in \mathds{U}$ requests the cloud infrastructure for execution of their respective applications. The workload processing information and multiple ($x$) resource utilization \{${R}es_1$, ${R}es_2$, ..., ${R}es_x$\} thereof are stored in a repository to provide training data for estimation of the forthcoming workload. A \textit{Resource Management Unit} (RMU) utilizes predicted workload information for resource provisioning, VM allocation and concerned management decisions. 

\par A Hadamard quantum mechanics based neural network is developed and employed for prediction of extensive range of cloud workloads. HQNN is a feed-forward deep neural network comprising of one input, multiple ($d$) hidden, and one output layers. It is represented as ${Y}={F}_{\Theta^\dagger}({X}^{\dagger})$, where $\Theta^\dagger$, ${X}^{\dagger}$, and ${Y}$ are network of qubit weights, qubit input vector, and the final output, respectively, wherein, the intended layers are inter-connected via qubit weights to build a quantum neural network.
The Qubit input vector \{$\Phi^I_1$, $\Phi^I_2$, ..., $\Phi^I_N$\}$\in {X}^{\dagger}$ obtained by transformation of $N$ previous and live resource usage samples \{${D}_1$, ${D}_2$, ..., ${D}_N$\} $\in {D}$  into Qubits, is  fed as a training data into the input layer of HQNN (Fig. \ref{fig:proposedmodel}). The proposed prediction model extracts  critical behavioral patterns from actual workload samples i.e., qubit input vector and analyses $N$ previous workload values to forecast the approaching load at next ${N+1}^{th}$ time-interval as predicted output (${D}^{Pr}$). Specifically, the controlled-Hadamard gate (${H}$) operation is applied at qubit neurons of multiple hidden and output layers to induce superposition during activation at respective neurons for enhanced learning of useful information and furnish high prediction accuracy. We have configured two distinct functional variants of QB-HNN model including \textit{QB-HNN for single attribute prediction} (QB-HNN-SAP) and \textit{QB-HNN for multiple attribute prediction} (QB-HNN-MAP) discussed in detail in Section \ref{sap} and Section \ref{map}, respectively.

\subsection{QB-HNN-SAP: Single Attribute Prediction Model} \label{sap}

The QB-HNN-SAP model is a specialized variant of the proposed quantum neural network designed to predict a single resource attribute such as CPU utilization within a cloud environment. Structurally, the model consists of one input layer, multiple ($d$) hidden layers, and a single output layer. These layers contain ${Q}^I$, ${Q}^H$, and ${Q}^O$ qubit neurons, respectively. The total size of the qubit network is quantified by Eq. \eqref{size}:
\begin{equation}
\resizebox{0.43\textwidth}{!}{$
S({X}^{\dagger}) = (Q^I \times Q^H) + (Q^H \times Q^H \times d) + (Q^H \times Q^O)$}
\label{size}
\end{equation}

To initiate training, the real-valued input samples normalized to $[0, 1]$ are encoded into a qubit input vector {$\Phi^I_1$, $\Phi^I_2$, ..., $\Phi^I_n$} $\in {X}^{\dagger}$ using Rotation ($R_\theta$) and Phase-Shift ($P_\theta$) gates. The transformation is governed by Eqs. \eqref{qubit} and \eqref{phase}, where $\Phi_i = \frac{\pi}{2} \times D_i$ (Eq. \eqref{qubit}) and phase shift operation is applied as:
\begin{gather} \label{qubit}
\Phi_i = \frac{\pi}{2} \times {D}_i	\\
f(\Phi_i)=
\begin{cases}
\frac{1}{\sqrt{2}}(\ket{0}), & {If(\lambda=0)} \\
\frac{1}{\sqrt{2}}(e^{i\theta}\ket{1}), & {If(\lambda=1)} \label{phase}
\end{cases}
\end{gather}

The forward propagation of qubit input vector is carried out via weighted synaptic connections between qubit neurons of different layers using Eqs. (\ref{l1}), (\ref{l2}), and (\ref{l3}), where $n \in$ \{${Q}^{I}$, ${Q}^{H}$, ${Q}^{O}$\}; $\uplus$ is a linear function computed at each qubit neuron; and ${B}$ is a bias qubit vector. The values of qubit weights \{$\Phi_1$, $\Phi_2$, ..., $\Phi_n$\}  determines the impact of the input vector on the output vector of the neurons and decides the strength of synaptic inter-connections between qubit neurons.  
\begin{gather}
	\uplus = (\Phi_1 \times \Theta_1) + (\Phi_2 \times \Theta_2) + ... +(\Phi_n \times \Theta_n) \label{l1}\\
	{X}^{\dagger}\cdot{\Theta^\dagger}=(\Phi_1 \times \Theta_1) + (\Phi_2 \times \Theta_2) + ... +(\Phi_n \times \Theta_n) \label{l2}\\
	\uplus=	{X}^{\dagger}\cdot{\Theta^\dagger} + {B} \label{l3}
\end{gather}
\begin{definition}
	\textbf{Superposition}: An exclusive property of Qubits that allows them to exist in multiple states or positions, simultaneously. This property helps to enrich the diversity of the Qubit population leading to enhanced optimization, faster convergence, and significant speed-up of the intended machine learning process.   
\end{definition} 
 \begin{definition}
	\textbf{Hadamard Gate (${H}$)}: A quantum operational gate that brings qubit into a superposition state which is equivalent to a complex linear integration of the basic 0 or 1 state. Specifically, it induces an equal probability of existence as `0' and `1' state within the respective qubit. Fig. \ref{fig:hadamard} depicts  schematic representation and operation of the Hadamard gate.

 \begin{figure}[!htbp]
		\centering
		\includegraphics[width=0.95\linewidth]{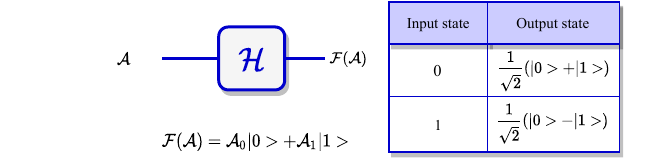}
		\caption{Hadamard Gate: symbolic and operational presentation}
		\label{fig:hadamard}
	\end{figure}	
\end{definition}

\begin{theorem}
 	Hadamard operation maps the input states to halfway between HIGH and LOW Logic levels.
 \end{theorem}
 
 	\begin{proof}
 		Classically, 
 		 Hadamard operation $H\equiv |\alpha\ket{0} + \beta\ket{1}| = \alpha \frac{\ket{0} + \ket{1}}{\sqrt{2}} + \beta \frac{\ket{0} - \ket{1}}{\sqrt{2}}$\\
 		The~logic~state:\\
 		$\ket{0}$ turns into $\frac{\ket{0} + \ket{1}}{\sqrt{2}}$ and \\
 		$\ket{1}$ turns into $\frac{\ket{0} - \ket{1}}{\sqrt{2}}$\\
 		Hence, LOW state is $1/\sqrt{2}$  times of $\ket{0}$ state shifted towards $\ket{1}$ state and\\
 		HIGH state is $1/\sqrt{2}$ times of $\ket{1}$ state shifted towards $\ket{0}$ state.
 	 Here $\alpha$ and $\beta$ are the complex numbers and $\ket{0}$ \& $\ket{1}$ are the quantum notations of computational basis states.
 	\end{proof}
\begin{corollary}
	Hadamard gate replicates the input for every even number of stages. 
\end{corollary}
\begin{proof}
	During Hadamard operation, \\
	$\ket{0}$ turns into $\frac{\ket{0} + \ket{1}}{\sqrt{2}}$ and \\
	$\ket{1}$ turns into $\frac{\ket{0} - \ket{1}}{\sqrt{2}}$\\
	In quantum algebra $H^2\equiv I$, which is an identity function\\
	Hence, $H^n$ maps to the same input logic if $n$ is even.
\end{proof}
  {The non-linearity element is incorporated at each neuron of multiple hidden  and  output layers by executing three consecutive steps: \textit{Qubit summation}, \textit{Superposition}, and \textit{Activation}. These steps stimulate the qubit network learning process with greater flexibility by creation of complex and intuitive patterns.} The Hidden layer neurons produce qubit vector \{$\Phi^{H_i}_1$, $\Phi^{H_i}_2$, ..., $\Phi^{H_i}_n$\} such that $1 \leq i \leq d$ by computing Eqs. (\ref{h1})-(\ref{h4}):
\begin{gather}
 \uplus_j^{H_i}=\sum_{i=1, j=1}^{d, {Q}^H} {{F}(\Phi_{ij}^{I}) \times {F}(\Theta_{ij}^{I}) } +{F}({B})
\label{h1}\\
\Phi_{j}^{H_i}={H}^{H_i}_j \otimes \uplus_j^{H_i}\label{h2} \\
{H}^{H_i}_j=\begin{cases}
\frac{1}{\sqrt{2}}(\ket{0}+\ket{1}), & {If(\lambda=0)} \\
\frac{1}{\sqrt{2}}(\ket{0}-\ket{1}), & {If(\lambda=1)} \label{h3}
\end{cases} \\	
{Y}_j^{H_i}={F}_{\Phi^I}(\Phi^{H_i}_j)  \label{h4}
\end{gather}
where, $\uplus_j^{H_i}$ represents qubit summation (Eq. (\ref{h1})) obtained for the  $j^{th}$ neuron of $i^{th}$ hidden layer. The adjusted qubit vector ($\Phi_j^{H_i}$) is obtained by applying superposition effect of Hadamard gate is applied to the qubit vector: $\uplus_j^{H_i}$ using Eqs. (\ref{h2}) and (\ref{h3}). 
 {The expression: ${F}(\Phi_j^{H_i})$ represents the  $j^{th}$ adjusted qubit vector obtained by applying activation function to $\Phi_j^{H_i}$ and Eq. (\ref{h4}) generates the final outcome ${Y}_j^{H_i}$  of $j^{th}$ neuron of the $i^{th}$ hidden layer.}

\par
 {At the final prediction stage, the qubit vector output \{$\Phi^{O}$\} is computed at the output layer by executing a series of quantum-inspired operations as formulated in Eqs. (\ref{o1})–(\ref{o4}). The process begins with the \textit{qubit-weighted summation} across all contributing hidden neurons (Eq. \ref{o1}), where each input from the hidden layer $\Phi_{ij}^{H}$ is modulated by its corresponding learned qubit weight $\Theta_{ij}^{H_d}$. This operation accumulates quantum-weighted influences, simulating the entangled contribution of qubit states in a superposed system.}

\begin{gather}
\uplus_j^{O}=\sum_{j=1}^{Q^O} F(\Phi_{ij}^{H}) \times F(\Theta_{ij}^{H_d}) \label{o1}
\end{gather}

\par
 {Subsequently, a \textit{controlled Hadamard transformation} is applied (Eq. \ref{o2}) to inject quantum superposition into the summation output $\uplus_j^{O}$. The transformation gate ${H}_j^O$ dynamically toggles between constructive and destructive superposition based on the control parameter $\lambda$ (Eq. \ref{o3}). This operation enriches the qubit signal $\Phi_j^{O}$ with non-linear quantum behavior, fostering better generalization across fluctuating workload patterns.}

\begin{gather}
\Phi_{j}^{O}={H}^{O}_j \otimes \uplus_j^{O} \label{o2} \\
{H}^{O}_j=
\begin{cases}
\frac{1}{\sqrt{2}}(\ket{0}+\ket{1}), & \text{if } \lambda=0 \\
\frac{1}{\sqrt{2}}(\ket{0}-\ket{1}), & \text{if } \lambda=1
\end{cases}
\label{o3}
\end{gather}

\par
Finally, the quantum-transformed signal $\Phi_j^{O}$ is passed through the \textit{quantum activation function} ${F}_{\Phi^I}(\cdot)$ to obtain the predicted output $Y_j^O$ (Eq. \ref{o4}). This function captures nonlinear mappings between qubit inputs and output space while maintaining quantum coherence.

\begin{gather}
{Y}_j^{O}={F}_{\Phi^I}(\Phi^{O}_j) \label{o4}
\end{gather}

\par
To ensure bounded output, a standard sigmoid activation function ${SIG}(\cdot)$ is used (Eq. \ref{sigmoidfunction}), effectively constraining the model's prediction in the range $[0, 1]$. This normalized range allows for direct comparability with actual workload traces during supervised training.

\begin{gather}
{SIG}(\varphi)=\frac{1}{1+e^{-\varphi}} \label{sigmoidfunction}
\end{gather}

\par
The model’s performance is quantitatively assessed using the Mean Squared Error (MSE) metric defined in Eq. (\ref{rmse}). This metric serves as the optimization fitness function during the QB-BiO learning process, ensuring convergence toward minimal prediction error across $m$ workload samples.

\begin{gather}
{E}val_{MSE} = \frac{1}{m} \sum_{i=1}^{m} ({D}^{Ac}_i - {D}^{Pr}_i)^2 \label{rmse}
\end{gather}

\subsection{QB-HNN-MAP: Multi-Attribute Prediction Model} \label{map}

To enable simultaneous forecasting of multiple cloud resource attributes (e.g., CPU, memory, and bandwidth), the QB-HNN model is extended to QB-HNN-MAP. Unlike conventional networks, QB-HNN-MAP embeds sets of qubit neurons in each layer, where the number of qubits in each set equals the number of attributes ($x$) to predict. Thus, input, hidden, and output layers consist of ${Q^\ast}^{I}$, ${Q^\ast}^{H}$, and ${Q^\ast}^{O}$ sets of qubit neurons, respectively. The network size is computed using Eq. \eqref{mimo1}:
\begin{equation}
\resizebox{0.43\textwidth}{!}{$ S^\ast(X^\dagger) = \sum_{i=1}^{x}\left( Q^{\ast I} \cdot Q^{\ast H} + Q^{\ast H} \cdot Q^{\ast H} \cdot d + Q^{\ast H} \cdot Q^{\ast O} \right) $}
\label{mimo1}
\end{equation}

The inter-layer synaptic connections are denoted by ${\Theta^\ast}^{IH}{ijk}$, ${\Theta^\ast}^{HH}{ijk}$, and ${\Theta^\ast}^{HO}_{ijk}$, representing connectivity between respective sets of qubit neurons. Hidden and output layer operations follow the same non-linear transformation principles as described earlier, utilizing Eqs. \eqref{h1}–\eqref{o4}.

QB-HNN-MAP supports joint prediction and classification by reusing shared qubit neural weights across attributes, enabling efficient resource provisioning decisions in multi-resource cloud environments.

\textit{Learning via QB-BiO}: Both QB-HNN-SAP and QB-HNN-MAP employ QB-BiO, a \textbf{Q}uantum \textbf{B}lackhole {Bi}phase {O}ptimization algorithm, for training. QB-BiO searches an initial population of $Z$ random qubit networks {$\Theta^\dagger_1$, $\Theta^\dagger_2$, ..., $\Theta^\dagger_Z$} to minimize the prediction error evaluated via $Eval()$. Through iterative exploration and exploitation, the algorithm selects optimal qubit weights and bias configurations. The training process occurs offline using prior workload data, while prediction occurs concurrently in real time. A detailed account of QB-BiO is presented in Section \ref{Qdpbho}.



\section{Qubit weights learning by QB-BiO } \label{Qdpbho}
\begin{figure}[!htbp]
    \centering
    \includegraphics[width=1.05\linewidth]{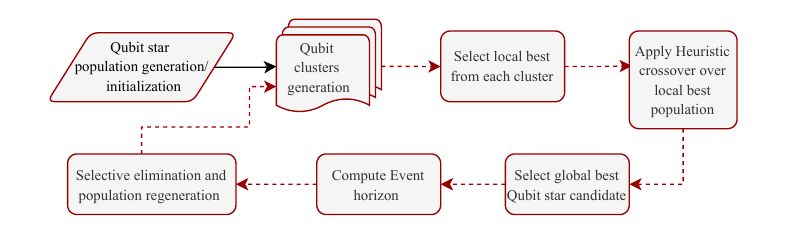}
    \caption{ {Simplified flowchart of QB-BiO optimization}}
    \label{fig:flowchart}
\end{figure}

The learning process of both QB-HNN-SAP and QB-HNN-MAP is driven by a novel Quantum BlackHole Biphase Optimization (QB-BiO) algorithm, which is inspired by the classical Black Hole Optimization (BHO) technique.  {BHO is a population-based metaheuristic that simulates the gravitational behavior of stars being attracted toward a central black hole, representing the best candidate solution \cite{hatamlou2013black}. In this context, QB-BiO iteratively updates the qubit-based synaptic weights \{$\Theta_1$, $\Theta_2$, ..., $\Theta_n$\} and the bias parameter (${B}$) by pulling them towards optimality in the solution space, such that the fitness function (${E}val_{MSE}$) is minimized, thereby enabling efficient quantum-inspired learning.} QB-BiO is a Qubit-population based  optimization algorithm which allows an extensive searching in multiple directions for selection of the most optimal solution. In QB-BiO, each candidate solution is considered as a Qubit star while a Qubit star having the best fitness value is designated as a `\textit{Quantum Black-hole}'. As illustrated in Fig. \ref{fig:proposedmodel}, the proposed optimization algorithm involves three consecutive stages: (\textit{i}) \textit{Qubit cluster optimization}, (\textit{ii}) \textit{Heuristic optimization}, and (\textit{iii}) \textit{Position Update}.  {Fig.~\ref{fig:flowchart} illustrates a simplified flowchart outlining the iterative and sequential steps involved in the QB-BiO optimization process.}

\subsection{Qubit cluster optimization}
In this stage, a random population of Qubit stars (i.e.,  Qubit network vectors) \{$\Theta^\dagger_1$, $\Theta^\dagger_2$, ..., $\Theta^\dagger_{Z}$\}$\in \Theta^\dagger$ is generated using Eq. (\ref{eqn:qq}), where $\alpha=\sqrt{(rd)}$ and $\beta=\sqrt{(1-rd)}$. The term $rd$ is a randomly generated number in range [0, 1], $\alpha$ and $\beta$ are the probability amplitudes for realizing $\ket{0}$ and $\ket{1}$ respectively, $\Psi$ is a quantum state in the Hilbert vector space and $\Theta$ is the corresponding qubit state.  The ${Z}$ Qubit stars are organized into ${K}$ Qubit clusters ($QC$) or sub-populations by using K-Means Clustering algorithm. The inter-clusters are distinguished on the basis of similarity of amplitude of the qubit weights within a network such that the Qubit clusters are  as different as possible. The qubit clustering is accomplished using Eq. (\ref{eq.cluster}), where ${W}_{ik}$ defines mapping of $i^{th}$ Qubit network ($\Theta^\dagger_i$) and $\mu_k$ is  the centroid of $k^{th}$ $QC$. 
\begin{gather}
\label{eqn:qq}
\Psi=\alpha + i\beta \\
\Theta= phase(\Psi)\times \frac{\pi}{2} 
\end{gather} 
\begin{equation}
\label{eq.cluster}
G=\sum_{i=1}^{{Z}}\sum_{k=1}^{{K}}{{W}_{ik}{|{\Theta^\dagger}_i- \mu_k|}^2}
\end{equation}

All the candidates of each qubit cluster ($\Theta^\dagger_{ik}: i \in [1, {Z}/{K}], k \in [1, {K}]$) are evaluated  using fitness evaluation function (Eq. (\ref{rmse})) over the sample data from repository. The best solution of each $k^{th}$ cluster  is observed as  blackhole (i.e., $QC_{k}^{best}$) such that $QC_{k}^{best} =$ Best(\{$\Theta^\dagger_1$, $\Theta^\dagger_2$, ..., $\Theta^\dagger_{{Z}/{K}}$\}). 
\subsection{Heuristic optimization}
The Qubit cluster blackholes ($QC_i^{best}$: $0 \leq i \leq {K}$) consitute the  population for next stage, wherein the diversity of the population is enhanced by executing a \textit{heuristic crossover} operation that produces new candidates of superior breed. During heuristic optimization, Qubit stars behave as chromosomes, wherein the two  parent chromosomes are randomly selected and their fitness values are evaluated using Eq. (\ref{rmse}) and compared to identify  the best parent chromosome.  Thereafter, Eq. (\ref{heuristic}) combines the two $QC^{best}$ blackholes based chromosomes to obtain a new offspring (i.e., Qubit network) which is more closer to the parent having better fitness value. Therefore, the diversity of the search space is improved significantly  by including new and better individuals in the successive optimization stage. 
Let $QC_{k}^{best}$ and $QC_{j}^{best}$ be the two parent chromosomes and $QC_{k}^{best}$ is observed as a best chromosome having greater fitness value. Consequently, a new offspring ($QG_k^{Off}$) is produced by employing Eq. (\ref{heuristic}):   
\begin{gather}
\label{heuristic}
QG_k^{Off} = {CO}_i \times (QC_{k}^{best}-QC_{j}^{best}) + QC_{k}^{best} 
\end{gather}
where, $k \neq j$, $i \in [1, {Z}/{K}]$, ${CO}_i$ is a crossover rate randomly generated in the range [0, 1] for $i^{th}$ gene. Likewise, ${K}$ new offsprings are produced i.e., one from each member of the population of Qubit cluster blackholes ($QC^{best}$). Eq. (\ref{g2}) selects the best between a new offspring ($QG_k^{Off}$)  and parent with lesser fitness ($QC_j^{best}$) to enhance the diversity of the Qubit cluster population by inducing Qubit network candidates with improved fitness values.

\begin{equation}\label{g2}
QC^{best}_{j}=\begin{cases}
QG_k^{Off} & \text{If (${F}$($QG_k$) $\geq$ ${F}$($QC^{best}_{j}$))} \\
QC^{best}_{j} & {\text{Otherwise}} 
\end{cases}
\end{equation}

Thereafter, a best among the members of heuristic population \{$QC_k$: $0 \leq k \leq {K}$\} is nominated as Qubit global blackhole ($QG^{best}$).

\subsection{Position Update}
The position of Qubit stars is changed with respect to the values of $QC_{k}^{best}$ and $QG^{best}$ using Eq. (\ref{bl1}), where ${\Theta^\dagger}_i^k(t)$ and ${\Theta^\dagger}_i^k(t+1)$ 
are the positions of $i^{th}$ star of $k^{th}$ sub population at time instance $t$ and $t+1$, respectively. $r_1$ and $r_2$ are random
numbers in the range [0, 1]  while $\alpha_c$
and $\alpha_h$ are the attraction forces applied on ${\Theta^\dagger}_i^k(t)$
by $QC_{k}^{best}$ and $QG^{best}$, respectively. Eq. (\ref{bl1}) employs
Qubit cluster blackhole ($QC_{k}^{best}$) and Qubit global blackhole ($QG^{best}$) for changing the positions of the Qubit stars to  maintain the diversity of Qubit star population  by  controlling the convergence speed gradually and retaining the desired level of exploration.

\begin{equation}\label{bl1}
\begin{aligned}
{CF}(t)= \alpha_c^k r_1\big({QC}_{k}^{best}(t)-{\Theta^\dagger}^i_k(t) \big)\\
{HF}(t)= \alpha_h r_2\big({QG}^{best}(t) - {\Theta^\dagger}^i_k(t)\big) \\
{\Theta^\dagger}_i^k(t+1)= {\Theta^\dagger}_i^k(t)+{CF}(t) +{HF}(t)
\end{aligned}
\end{equation}

\par All the updated qubit stars are evaluated using Eq. (\ref{rmse}) to locate a better solution (qubit network) within $k^{th}$ Qubit cluster. Accordingly, the values of $QC_k^{best}$ and $QG^{best}$ are relocated as per the admissibility. The proposed QB-BiO algorithm follows a natural blackhole phenomenon, wherein a blackhole engulfs all the things that enter into it including light. In that event, QB-BiO prohibits the entry of the candidate solution returning back from an event horizon (${EH}$) area of a blackhole solution. ${EH}$ is marked-out by the radius of event horizon (${R}_{{EH}}$). The ratio of the fitness value of a Qubit cluster blackhole (${F}(QC^{best})$) and  summation of the fitness values of its sub-population ($\sum_{i=1}^{{Z}/{K}}{{F}({{\Theta^\dagger}^i_k})}$) enumerates an event horizon radius (${R}_{{EH}}$) of the respective blackhole using Eq. (\ref{r1}). Similarly, the event horizon radius of a global blackhole (${R}_{{EH}}\big(QG_{k}^{best}\big)$)  is evaluated using Eq. (\ref{r2}), where ${F}(QG_{k}^{best})$ is fitness value of  heuristic global blackhole, $\sum_{k=1}^{K}\sum_{i=1}^{{Z}/{K}}{{F}({\Theta^\dagger}_{k}^{i})}$ is a fitness value of the entire population.

\begin{equation} \label{r1}
{R}_{{EH}}\big(QC_{k}^{best}\big)=\frac{{F}(QC_{k}^{best})}{\sum_{i=1}^{{Z}/{K}}{{F}({\Theta^\dagger}_{k}^{i})}} \quad k \in [1, K]
\end{equation}

\begin{equation} \label{r2}
{R}_{{EH}}\big(QG^{best}\big)=\frac{{F}(QG_{k}^{best})}{\sum_{k=1}^{{K}}\sum_{i=1}^{{Z}/ {K}}{{F}({\Theta^\dagger}_{k}^{i})}} 
\end{equation}

The distance between the best and rest of the solutions is determined by evaluating
the arithmetic difference of their fitness values to confirm that a member solution
has entered into an event horizon of the blackhole solution. Each Qubit star attracts towards the Qubit cluster and  Heuristic global blackholes and their corresponding distances from  these two blackholes are computed  using  Eqs. (\ref{d1}) and  (\ref{d2}):
\begin{gather}
\resizebox{0.43\textwidth}{!}{$ 
{D}is_{QC_{k}^{best}}\big(\Theta_{k}^{i}\big)={F}(QC_{k}^{best}) - {F}(\Theta_{k}^{i}) \quad i \in [1, {Z/K}]  $}\label{d1}
\\ \label{d2}
{D}is_{QG_{k}^{best}}\big(\Theta_{k}^{i}\big)={F}(QG_{k}^{best}) - {F}(\Theta_{k}^{i}) \quad i \in [1, 2{K}]
\end{gather}
wherein ${D}is_{QC_{k}^{best}}\big(\Theta_{k}^{i}\big)$ and ${D}is_{QG_{k}^{best}}\big(\Theta_{k}^{i}\big)$ are the distances of $i^{th}$ Qubit star ($\Theta_{k}^{i}$) of $k^{th}$ cluster from Qubit cluster blackhole ($QC_{k}^{best}$) and global blackhole ($QG^{best}$), respectively. 
\begin{theorem}
 	Given ${F}(\Theta_{k}^{i})$, ${F}(QC_{k}^{best})$, and ${F}(QG^{best})$ are the fitness values of $i^{th}$ Qubit star ($\Theta_{k}^{i}$) of $k^{th}$ cluster, blackhole of $k^{th}$ cluster, and  Heuristic global blackhole, respectively. The sustainability of $\Theta_{k}^{i}$ is bounded by the distances: ${D}is_{QC_{k}^{best}}\big(\Theta_{k}^{i}\big)$ (Eq. (\ref{d1})) and ${D}is_{QG_{k}^{best}}\big(\Theta_{k}^{i}\big)$ (Eq. (\ref{d2})).\\
	For the event horizon radii: ${R}_{{EH}}\big(QC_{k}^{best}\big)$ and ${R}_{{EH}}\big(QG^{best}\big)$ of cluster blackhole and global blackhole, respectively; 
	\begin{itemize}
		\item[(i)] If $\big({D}is_{QC_{k}^{best}}\big(\Theta_{k}^{i}\big) \leq $
		${R}_{{EH}}\big(QC_{k}^{best}\big)\big)$ $\vee$ \big(${D}is_{QG^{best}}\big(\Theta_{k}^{i}\big) \leq $
		${R}_{{EH}}\big(QG^{best}\big)$\big), then $\Theta_{k}^{i}$ is collapsed.
		\item[(ii)] If $\big({F}(\Theta_{k}^{i}) \approxeq {F}(QC_{k}^{best})\big) \vee \big({F}(\Theta_{k}^{i}) \approxeq {F}(QG^{best})\big)$, then $\Theta_{k}^{i}$ is replaced by $\Theta_{k}^{New}$ to retain the  uniform number of solutions and diversity of the Qubit population.
		\end{itemize}
	\end{theorem}

\begin{proof}
Let ${F}(\Theta_{k}^{1})$, ${F}(\Theta_{k}^{2})$, ..., ${F}(\Theta_{k}^{n})$ are the fitness values of $n$ Qubit stars within $k^{th}$ cluster\\
Then, ${F}(QC_{k}^{best})=$ MIN(${F}(\Theta_{k}^{1})$, ${F}(\Theta_{k}^{2})$, ..., ${F}(\Theta_{k}^{n})$)\\
$\sum_{i=1}^{n}{{F}(\Theta_{k})}=$ ${F}(\Theta_{k}^{1})$+ ${F}(\Theta_{k}^{2})$+ ...+ ${F}(\Theta_{k}^{n})$
$$ {R}_{{EH}}\big(QC_{k}^{best}\big)=\frac{{F}(QC_{k}^{best})}{\sum_{i=1}^{{Z}/{K}}{{F}({\Theta^\dagger}_{k}^{i})}} \quad \text{(Using Eq. (\ref{r1}))}$$ 
Since prediction error constitutes fitness evaluation function (Eq. (\ref{rmse})), it is intuitively minimized.\\ Therefore, $0 <{R}_{{EH}}\big(QC_{k}^{best}\big) \leq 1$ (Step 1) by reason of  ${F}(QC_{k}^{best}) \leq  \sum_{i=1}^{n}{{F}(\Theta_{k})}$. 
$$ {D}is_{QC_{k}^{best}}\big(\Theta_{k}^{i}\big)=|{F}(QC_{k}^{best}) - {F}(\Theta_{k}^{i})|\quad \text{(Using Eq. (\ref{d1}))} $$ 
$\Rightarrow 0 \leq {D}is_{QC_{k}^{best}}\big(\Theta_{k}^{i}\big) \leq {F}(\Theta_{k}^{i}) < 1 $ (Step 2)\\
\begin{itemize}
\item[(i)] From Step 1 and 2, the possible range of values for ${R}_{{EH}}\big(QC_{k}^{best}\big) \in (0, 1]$ and ${D}is_{QC_{k}^{best}}\big(\Theta_{k}^{i}\big) \in [0, 1) $\\
When the condition: ${D}is_{QC_{k}^{best}}\big(\Theta_{k}^{i}\big) \leq$ ${R}_{{EH}}\big(QC_{k}^{best}\big)$ is TRUE, the Qubit star (${F}(\Theta_{k}^{i})$) enters into the event horizon of a cluster blackhole. $\quad$ (Step 3)	   
\\ Following the same procedure for global blackhole, when the ${F}(\Theta_{k}^{i})$ enters into the event horizon radius of the global blackhole (${R}_{{EH}}\big(QG^{best}\big)$)  $\quad$ (Step 4) 

Therefore, if either Step 3 or Step 4 is TRUE, the  Qubit star (${F}(\Theta_{k}^{i})$) collapses forever.
 \item[(ii)] If either $\big({F}(\Theta_{k}^{i}) \approxeq {F}(QC_{k}^{best})\big)$ or $\big({F}(\Theta_{k}^{i}) \approxeq {F}(QG^{best})\big)$ is TRUE, then ${D}is_{QC_{k}^{best}}\big(\Theta_{k}^{i}\big) \approx 0$ OR ${D}is_{QG^{best}}\big(\Theta_{k}^{i}\big) \approx 0$ becomes TRUE\\  $\Rightarrow \big({D}is_{QC_{k}^{best}}\big(\Theta_{k}^{i}\big) \leq $
 ${R}_{{EH}}\big(QC_{k}^{best}\big)\big)$ $\vee$ \big(${D}is_{QG^{best}}\big(\Theta_{k}^{i}\big) \leq $
 ${R}_{{EH}}\big(QG^{best}\big)$\big) become TRUE which in turn stimulate replacement of $\Theta_{k}^{i}$ by introduction of $\Theta_{k}^{New}$ with varying fitness value and high diversity.
\end{itemize}
\end{proof}
 The operational summary of QB-BiO is depicted in Algorithm \ref{algo-dpbho}. \textit{Time Complexity}: Step 1 initializes random solutions, has complexity ${O}(1)$. Step 2 trains HQNN and evaluates ${M}$ samples for fitness of ${Z}$ solutions across network: $({{Q}^I} \times {{Q}^H}) + ({{Q}^H} \times {{Q}^I})$, wherein ${{Q}^I} > {{Q}^H}$ and  ${{Q}^O}=1$ producing ${O}({Z} \times {{Q}^I}^2 \times {M})$ complexity. Steps (3-5), steps (6-12), and steps (13-15) iterate ${K}$ times have equal time complexity of ${O}({K})$. Assume steps (16-29) repeat for $t$ intervals, wherein steps (19-21) have ${O}({K})$  while steps (24-28) have ${O}({Z})$ complexities. Hence, the total time complexity for QB-BiO algorithm is ${O}({Z}\times {{Q}^I}^2 \times {M} \times {K})$.  
\begin{figure}[!htbp]
	\removelatexerror
	\begin{algorithm}[H]
		\caption{HQNN Optimization by QB-BiO}
		\label{algo-dpbho}
		{
			Initialize $N$ random solutions: \{${\Theta^\dagger}_1$, ${\Theta^\dagger}_2$, ..., ${\Theta^\dagger}_{Z}$\}$\in {X}$ \;	
			Organize ${X}$ into ${K}$ Qubit clusters and evaluate each random solution on training data using Eq. (\ref{rmse}) \;
			\For{each k = \{1, 2, ..., ${K}$\}}{$QC_{k}^{best}$ = Best(\{${\Theta^\dagger}_1$, ${\Theta^\dagger}_2$, ..., ${\Theta^\dagger}_{{Z}/ {K}}$\})\; }
			\For{ i= \{1, 2, ..., ${K}$\}}{Two parents ($QC_{k}^{best}$ and $QC_{j}^{best}$) are randomly selected and a new offspring ($QG_{k}^{Off}$) is produced using Eq. (\ref{heuristic})\;
				Evaluate $QG^{Off}_{k}$ over training data using Eq. (\ref{rmse}) \;
				\If{fitness value of $QG^{Off}_{k} \geq$  less best parent ($QC_{j}^{best}$) }{
					Replace  $QC_{j}^{best}$ with $QG^{Off}_{k}$ \; 
				}
			}
			\For{each k = \{1, 2, ..., ${K}$\}}{$QG^{best}$ = Best(\{$QC_1^{best}$, $QC_2^{best}$, ..., $QC_{{K}}^{best}$\}) \; }
			\While{termination }{
				Update position of ${\Theta^\dagger}^i_k$ using Eq. (\ref{bl1})\;
				Evaluate ${\Theta^\dagger}^i_k (t+1)$ using Eq. (\ref{rmse}) \;
				\For{each k = \{1, 2, ..., ${K}$\}}{$QC_{k}^{best}(t+1)$ = Best($QC_{k}^{best}(t)$, \{${\Theta^\dagger}_1$, ${\Theta^\dagger}_2$, ..., ${\Theta^\dagger}_{{Z}/ {K}}$\})\; 
				}
				$QG_{best}(t+1)$ = Best($QG_{best}(t)$, \{$QC_1^{best}$, $QC_2^{best}$, ..., $QC_{{K}}^{best}$\}) \;	
				Estimate the radius and distances using Eqs. (\ref{r1}-\ref{d2}) \;
				\For{each i = \{1, 2, ..., ${Z}$\}}{
					\If{(${D}is_{QC_{k}^{best}}\big(\Theta_{k}^{i}\big) \leq $
						${R}_{{EH}}\big(QC_{k}^{best}\big)$) $\vee$ (${D}is_{QG^{best}}\big(\Theta_{k}^{i}\big) \leq $
						${R}_{{EH}}\big(QG^{best}\big)$) }{Collapse ${\Theta^\dagger}^i_k$ and rejuvenate population by adding a new candidate using Eq. \ref{eqn:qq}\;}

				}		
			}		
	} 	\end{algorithm}
	
\end{figure}

{ \subsection{Convergence and Stability Analysis of QB-BiO}

\begin{theorem}[Convergence and Stability of QB-BiO]
Let $F(\Theta)$ be a bounded and continuous fitness function over the search space $\Omega$, and assume that the initial population of qubit stars $\{\Theta^\dagger_1, \Theta^\dagger_2, ..., \Theta^\dagger_Z\} \in \Omega$ is diverse and randomly initialized. Then, the QB-BiO algorithm converges ``with probability 1'' (almost surely) to a global or near-global optimum of $F(\Theta)$ under the following conditions:
\begin{itemize}
    \item[(i)] The crossover rate ${CO}_i$ is uniformly sampled in $[0,1]$ at each generation, ensuring exploration.
    \item[(ii)] The attraction forces $\alpha_c, \alpha_h \in (0,1]$ are non-zero and gradually decreased to favor exploitation.
    \item[(iii)] The fitness landscape of $F(\Theta)$ has a finite number of local minima and a global minimum.
    \item[(iv)] Event horizon radii ${R}_{EH}$ are updated based on relative fitness, such that weaker solutions are probabilistically eliminated and replaced with diverse candidates.
\end{itemize}
Under these assumptions, the QB-BiO algorithm is both:
\begin{itemize}
    \item[(a)] \textit{Convergent}: It asymptotically guides the population towards higher fitness regions with probability 1 as the number of generations $t \to \infty$.
    \item[(b)] \textit{Stable}: The variance of fitness values in the population decreases over time while preserving population diversity via the event horizon mechanism.
\end{itemize}
\end{theorem}

\begin{proof}
We analyze convergence using properties of stochastic search and Markov chains.

\textit{Step 1: Markov Property and Population Coverage.} \\
The evolution of the population in QB-BiO depends only on its current state and not on its history due to the use of random sampling ($r_1$, $r_2$, $CO_i$) and selection. Therefore, the QB-BiO process can be modeled as a time-inhomogeneous Markov chain over a finite population space $\Omega^Z$.

\textit{Step 2: Global Search Property (Ergodicity).} \\
Due to random initialization and stochastic position updates (Eqs. \ref{bl1}, \ref{heuristic}), every region of the search space $\Omega$ can be visited with non-zero probability. The event horizon-based elimination and re-initialization mechanism ensures that diversity is maintained in each iteration. Hence, the Markov chain induced by QB-BiO is irreducible and aperiodic.

\textit{Step 3: Selection Pressure and Convergence.} \\
The fitness-based selection in each cluster and globally ensures monotonic improvement in the best fitness $F(QG^{best})$ over time, or at least no degradation. This is due to the greedy update in Eq. (\ref{g2}) and fitness-based replacement post event horizon collapse. Hence, $\lim_{t\rightarrow\infty} F(QG^{best}_t) = F^*$ exists, where $F^*$ is a local or global optimum. If $F$ is unimodal or multimodal with isolated minima, convergence to global optimum occurs with probability 1 due to ergodicity.

\textit{Step 4: Stability Analysis.} \\
Let $\sigma^2_t$ denote the variance of fitness values in the population at iteration $t$. As weaker solutions collapse into the blackhole (Eq. \ref{d1}-\ref{d2}) and are replaced by new diverse candidates, $\sigma^2_t$ does not explode but tends to decrease while maintaining bounded diversity. The attraction terms $\alpha_c$ and $\alpha_h$ are assumed to decay over time, enforcing a transition from exploration to exploitation. Thus, $\lim_{t \to \infty} \sigma^2_t \to 0$ implies that all surviving solutions cluster around the best fitness region.

\textit{Step 5: Almost Sure Convergence.} \\
From the ergodicity of the Markov process, the boundedness of the fitness landscape, and the greedy fitness-based selection, QB-BiO converges with probability 1 to a region arbitrarily close to the global optimum, i.e.,
\[
\lim_{t \to \infty} \Pr \left( \| \Theta_t - \Theta^* \| < \epsilon \right) = 1, \quad \forall \epsilon > 0
\]
where $\Theta^*$ denotes the global optimum in $\Omega$.

\end{proof}
}
\section{Performance Evaluation }

\subsection{Experimental Set-up}
The simulation experiments are conducted on a server machine assembled with two Intel\textsuperscript{\textregistered} Xeon\textsuperscript{\textregistered} Silver 4114 CPUs with a 40 core processor and a 2.20 GHz clock speed. The computation machine is deployed with 64-bit Ubuntu 16.04 LTS, having 128 GB RAM, and the proposed model is implemented in Python 3.7.  {Table \ref{table:name1} presents the experimental configuration of the proposed QB-HNN model, which outlines the architecture of the neural network, the training ratio and the quantum optimization settings. The size of the hidden layer (${Q}^H$ = 7) was determined by extensive empirical tuning to achieve an optimal balance between model complexity and convergence efficiency. Using the superposition property of Hadamard qubits, each neuron represents multiple quantum states, enabling rich feature learning with fewer nodes. Uniform hyperparameters were maintained across all datasets to assess generalization, while the QB-BiO algorithm adaptively fine-tuned the qubit weights for each workload.}

\begin{table}[!htbp]
	\centering
	
	\caption[Table caption text] {Experimental set-up parameters and values}  
	\label{table:name1}
	\begin{tabular}{l c}
		\hline
		\textbf{Parameter }   &\textbf{Value}  \\
		\hline
		Input neural nodes (${Q}^I$)     & 10 \\ 	\hline
		Hidden layer nodes (${Q}^H$)        & 7       \\ 	\hline
		Output layer nodes(${Q}^O$)      & 1          \\ 	\hline
		Maximum epochs($G_{max}$)     & 50        \\ 	\hline
		 {ratio of training data} & 75\%            \\ 	\hline
		Size of Qubit star population  & 15-36\\ 	\hline
		Number of Qubit clusters & 4 \\
		\hline
	\end{tabular}
\end{table}

\subsection{Datasets}

 {To evaluate the effectiveness and robustness of the proposed QB-HNN model in multivariate time-series forecasting, six benchmark datasets spanning three workload categories were employed: \textit{Cluster workloads}, \textit{Web server workloads}, and \textit{High Performance Computing (HPC) workloads}. The cluster workloads are extracted from the Google Cluster Data (GCD) traces, which consist of fine-grained CPU and memory usage logs recorded every 5 minutes over a 29-day period. For analysis, we aggregated resource metrics during the first 10 days, resulting in 2 million job entries each for CPU ($G^C$) and memory ($G^M$) usage, with two channels (CPU and memory). Prediction windows of 5, 10, 30, and 60 minutes were used to generate temporal segments with corresponding visible lengths of 60 minutes. Each input sample was structured as a multivariate time series with sliding windows to simulate real-time forecasting conditions.}

 {The web server workloads include NASA-HTTP ($N$) and Saskatchewan ($S$) web access traces spanning 31 days and 7 months, respectively. These contain time-stamped HTTP request and response logs, which were aggregated into 5- to 60 minute intervals and normalized to create univariate time series for traffic volume prediction. The HPC workloads, AuverGrid ($Au$) and SHARCNet ($Sh$) cover 365 and 11 days of job submissions, respectively, with millions of job records and high variability in job execution time. The input vectors for all HPC datasets were created using job submission time-series aggregated by window size. Table~\ref{table:(NASA)workloadcharacteristic} summarizes key statistical properties of each dataset, including mean and standard deviation of resource usage. This diverse dataset composition poses challenges in terms of scale, volatility, and workload heterogeneity, providing a comprehensive testbed for evaluating the generalization capability of QB-HNN across varied cloud environments.}
 Table \ref{table:(NASA)workloadcharacteristic} shows the statistical characteristics of the evaluated workloads. 

\begin{table}[!htbp] 	
	\caption[Table caption text] {Characteristics of evaluated workloads  }  
	\label{table:(NASA)workloadcharacteristic} 	
	\centering
	\resizebox{9cm}{!}{
		\begin{tabular}{  l  l  l  l   l  l  }
			\hline
			\textbf{Category}	&\textbf{ Workload} &\textbf{ Duration} & \textbf{ Jobs}& \textbf{Mean(\%)} & \textbf{St.dev} \\
			
			\hline
			\multirow{3}{*}{\textbf{Cluster }} &GCD-CPU ($G^C$) & 10 days&2 M &21.84&13.62\\
			&GCD-memory ($G^M$) & 10 days& 2 M& 19.55&16.6 \\
			
			\hline
			\multirow{2}{*}{\textbf{Web }}&	NASA-HTTP ($N$)& 31 days& 1.8 M& 2.95&1.72\\
			&{Saskatchewan} ($S$) &7 months&2.5 M &  2.97&1.73 \\ 					
			
			\hline
			\multirow{3}{*}{\textbf{HPC}} &AuverGrid ($Au$) & 365 days& 2.3 M &1.27E+9 &4.18E+4 \\	
			&SHARCNet ($Sh$)&11 days&188K&1.14E+9  &4.89E+5 \\
			\hline
	\end{tabular}}
\end{table}

\subsection{Results}
\begin{figure*}[!htbp]
	\centering	
	\subfigure[Cluster workload: GCD-CPU ]{\includegraphics[width=0.31\linewidth, scale=3]{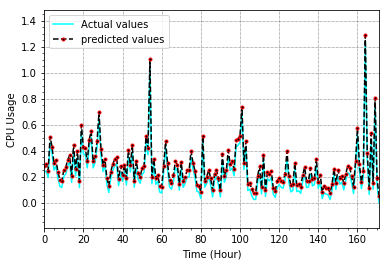}\label{ru1}} 
	\subfigure[Web workload: NASA]{\includegraphics[width=0.31\linewidth, scale=3]{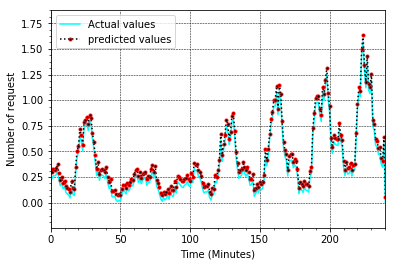}\label{ru3}} 
	\subfigure[HPC Grid workload: SHARCNet]{\includegraphics[width=0.31\linewidth, scale=2]{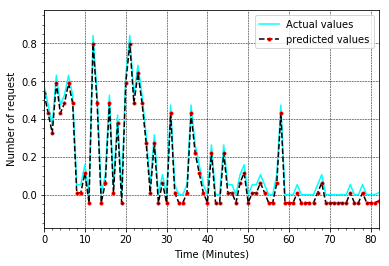}\label{GCDMin}} 	
	\caption{Actual v/s Predicted values}
	\label{gcd}
	
\end{figure*}

The performance of QB-HNN prediction model is thoroughly investigated using extensive range of heterogeneous cloud applications and variable resource utilization by VMs. Fig. \ref{gcd} compares actual and predicted values of three different categories of workloads including GCD-CPU cluster (Fig. \ref{gcd}a), NASA web traces (Fig. \ref{gcd}b), and SHARCNet HPC Grid (Fig. \ref{gcd}c) workloads. It is observed from the figures that the proposed method estimate the future values  closer to the actual values for all different categories of workloads. Moreover, the sudden upsurge (sudden peaks) and plunge (random fall) of the different workload demands are precisely predicted by the proposed method.

Tables \ref{table:resultsweb} and Table \ref{table:results} show the detailed analysis of various performance metrics related to the evaluation of QB-HNN-SAP and QB-HNN-MAP, respectively, for all workload categories.  {The performance metrics reported in this study: MAE and MSE are standard and are used according to their conventional definitions.} The prediction error values including mean square error (MSE) and mean absolute error (MAE) obtained for smaller prediction-interval (i.e., 5 minutes) is lesser than that for larger PWS (i.e., 60 minutes) because of the learning of QB-HNN with a greater number of training and retraining data samples (as obtained during data aggregation) for PWS of 5 minutes as compared to that of 60 minutes.  Among all the three categories of datasets, the values of MSE and MAE are highest for HPC Grid workloads because of the availability of smaller number of data samples as compared with rest of the datasets. Moreover, it is noticed that the average training time (ATT) and average memory space consumption (AMC) is more for 5 minutes prediction intervals because of the intended training process of QB-HNN with larger number of Qubit data values that consume enough time as well as memory space (in bytes). The values obtained for convergence during training of QHNN with QB-BiO  algorithm is lesser than 20 epochs for all variety of cloud workloads. This is due to the faster but not-premature convergence capability of proposed QB-BiO  algorithm which furnishes the learning process of QHNN in lesser number of iterations.

\begin{table}[!htbp] 	
	\caption[Table caption text] {Performance metrics for QB-HNN-SAP   }  
	\label{table:resultsweb} 	
	\centering
	\resizebox{8.7cm}{!}{
		\begin{tabular}{  c  c  c  c   c c c c }
			\hline
		
		{\textbf{Category}}	& {\textbf{ DS}$^a$}& {\textbf{ PWS}$^b$} & \textbf{ MSE}& \textbf{MAE} & \textbf{ATT}$^c$ & \textbf{Epochs} &\textbf{AMC}$^d$ 
			\\
			
			\hline
			\multirow{6}{*}[-0.4ex]{\rotatebox{90}{\textbf{Cluster}}}&\multirow{3}{*}{{$G^C$}} &5 &8.3E-4 & 0.034 &213 &15&2.85E+5 \\
		&	&30   & 9.4E-4& 0.059& 10.6&13&2.39E+5 \\
			
		&	&60  & 1.1E-3& 0.067& 8.30&15& 1.94E+5\\ \cline{2-8}
		&	\multirow{3}{*}{{$G^M$ }} &5 & 9.1E-4&0.047&246&13&2.44E+5\\		
			
		&	&30 & 9.5E-4& 0.061& 19.5&14&1.99E+5 \\
			
		&	&60  & 1.2E-3& 0.079& 11.0&16&1.29E+5\\ \hline	
			
			\multirow{6}{*}[-0.4ex]{\rotatebox{90}{\textbf{Web }}}&	\multirow{3}{*}{{$N$ }} &5 & 9.1E-4&0.085 &316&13&3.12E+5\\
			&	&30  & 1.3E-3& 0.099& 35.7& 14&2.94E+5 \\
			
			&	&60  & 2.2E-3& 0.107& 30.0& 17&2.45E+5 \\ \cline{2-8}
			&	\multirow{3}{*}{{$S$ }} &5 & 7.1E-4& 0.006& 294 &18 & 3.11E+5 \\		
			
			&	&30  & 1.1E-3& 0.018& 27.8 &16 & 2.72E+5\\
			
			&	&60 & 5.3E-3& 0.017& 20.3 &15 &2.14E+5 \\ \hline
			
			\multirow{6}{*}[-0.4ex]{\rotatebox{90}{\textbf{HPC Grid }}}&	\multirow{3}{*}{{$Au$  }} &5 &  1.1E-3& 0.107& 89.1&13&3.15E+5\\
			&	&30  &6.4E-3& 0.189& 7.35&17&2.43E+5 \\
			
			&	&60& 9.1E-3& 0.201& 5.48&15&1.75E+5\\ \cline{2-8}
			&	\multirow{3}{*}{{$Sh$ }} &5 & 5.8E-3& 0.119& 91.7&17&2.75E+5\\		
			
			&	&30 &8.7E-3& 0.149& 8.41&16&2.38E+5 \\
			
			&	&60  & 9.9E-3& 0.191& 6.73&14&1.98E+5 \\ \hline	
			
	\end{tabular}}
	\footnotesize{\tiny{{$^a$ DS: Datasets, $^b$ PWS: Prediction Window Size, $^c$ ATT: Average Training-Time in seconds, $^d$ AMC: Average memory space consumed (Bytes)}}}	
\end{table}
The evaluation of QB-HNN-MAP is performed using Cluster workloads (as reported in Table \ref{table:results}) which provides information of multiple resource such as CPU and memory usage while the other two categories of datasets consists of single attribute information only.  {The QB-HNN-MAP model was evaluated using the Google Cluster workloads (GC and GM), which contain synchronized multivariate resource attributes (CPU and memory). Other datasets, such as NASA-HTTP, Saskatchewan, AuverGrid, and SHARCNet, are univariate and therefore evaluated using QB-HNN-SAP.} The prediction error values obtained for QB-HNN-MAP are acceptable as compared with the values of MSE and MAE produced by QB-HNN-SAP with consumption of similar number of epochs during optimization. The accuracy of predicted values obtained using QB-HNN-MAP are equivalent to that of the prediction accuracy of the QB-HNN-SAP at the cost of increased ATT and AMC.
\begin{table}[!htbp] 	
	\caption[Table caption text] {Performance metrics for QB-HNN-MAP  }  
	\label{table:results} 	
	\centering
	\resizebox{8.7cm}{!}{
		\begin{tabular}{ c   c  c  c   c c c c   }
			\hline
			{\textbf{Category}}&	{\textbf{PWS}}&{\textbf{DS}} & \textbf{ MSE}& \textbf{MAE} & \textbf{ATT} & \textbf{Epochs} &\textbf{AMC} 
			\\ \cline{1-3} \cline{3-8}  
			
			\multirow{6}{*}[-0.4ex]{\rotatebox{90}{\textbf{Cluster}}}&	\multirow{2}{*}{5} &	{{$G^C$ }} &9.9E-4 & 0.056 &\multirow{2}{*}{232} &\multirow{2}{*}{15}&\multirow{2}{*}{4.85E+5}\\
			&	& {{$G^M$ }}	   & 9.4E-4& 0.059& & & \\ \cline{2-8} 
			
			&	\multirow{2}{*}{30} &	{{$G^C$}} &1.1E-3 & 0.094 &\multirow{2}{*}{218} &\multirow{2}{*}{15}&\multirow{2}{*}{4.17E+5}\\
			&	& {{$G^M$ }}	   & 1.2E-3& 0.059& & & \\ \cline{2-8} 
			
			&	\multirow{2}{*}{60} &	{{$G^C$ }} &1.9E-3 & 0.103 &\multirow{2}{*}{204} &\multirow{2}{*}{13}&\multirow{2}{*}{4.02E+5}\\
			&	& {{$G^M$ }}	   & 1.7E-3& 0.101& & & \\ \hline		
			
	\end{tabular}}
	
\end{table}
 It is observed that the average of training time for simultaneous prediction of multiple resources is longer and memory space consumption is almost two times of the AMC applied for single resource prediction. This is because of the presence of twice number of data values and variables within the main memory which require longer and sufficient time  for learning process. Similar to QB-HNN-SAP, it has been analysed that the consumption of memory space and time is reduced with the increasing size of the prediction interval for QB-HNN-MAP because of the usage of lesser number of aggregated and normalized data samples for training. As a consequence, it is concluded that QB-HNN-MAP is more applicable for simultaneous prediction of multiple attributes for the higher prediction interval or with smaller ratio of training dataset.

\subsection{Comparison}

The proposed prediction model is compared with widely used Neural Network (NN) based on Backpropagation (BPNN) \cite{lu2016rvlbpnn}, neural network optimized with Self-adaptive Differential Evolution (SaDE) algorithm \cite{kumar2018workload}, neural network optimized with Biphase-adaptive Differential Evolution (BaDE) algorithm \cite{kumar2018workload}, LSTM-RNN \cite{kumar2018long}, Deep Learning \cite{zhang2018efficient}, and Evolutionary Quantum Neural Network (EQNN) \cite{singh2021quantum} workload prediction models. The concise description of these approaches is given in the related work (Section \ref{rw}). The comparison is analysed in terms of \textit{Normalised MSE}, \textit{Normalised MAE},  \textit{Absolute Error Frequency (AEF)}, \textit{Prediction accuracy}, \textit{ATT}, \textit{Computational complexity} and \textit{Convergence} of training algorithm.

\subsubsection{Normalised MSE}
\begin{figure*}[!htbp]
	\centering	
	\subfigure[Web traces]{\includegraphics[width=0.31\linewidth, scale=3]{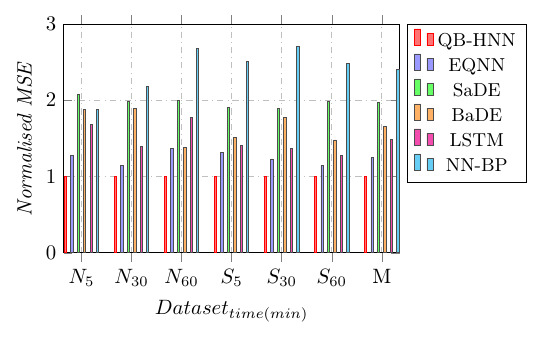}\label{ru1}} 
	\subfigure[Google Cluster traces]{\includegraphics[width=0.31\linewidth, scale=3]{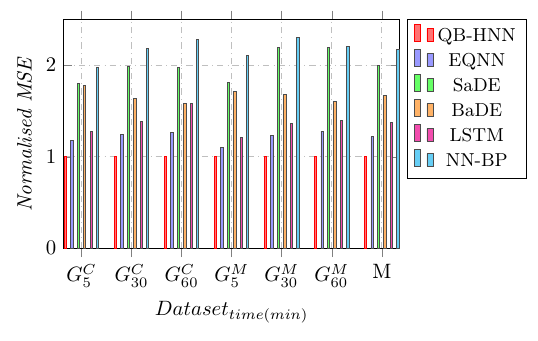}\label{ru3}} 
	\subfigure[High Performance/Grid traces]{\includegraphics[width=0.31\linewidth, scale=2]{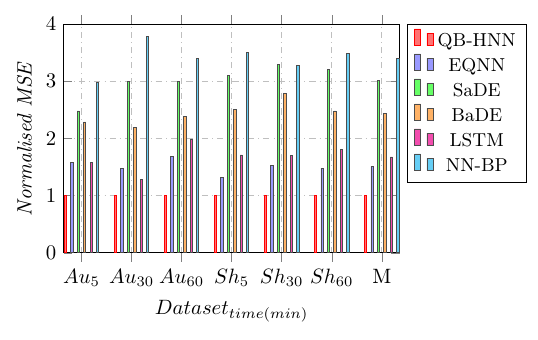}\label{GCDMin}} 	
	\caption{Normalised MSE}
	\label{mse}	
\end{figure*}

 {Fig. \ref{mse}} compares the normalised MSE results  obtained for QB-HNN with the relative MSE of five comparative approaches for all the three categories of dynamic cloud workloads. The value $1.00$ indicates the outcome of the proposed QB-HNN prediction model and greater values indicate worse performance. Therefore, all the MSE results obtained by applying different existing prediction methods are normalized with respect to MSE of QB-HNN. The error is reduced by an average of 22.83\%, 72.15\%, 72.16\%, 36.36\%, 75.12\% over EQNN, SaDE, BaDE, LSTM, and  NN-BP, respectively. The prediction error decreases in the order NN-BP $>$ SaDE $>$ BaDE $>$ LSTM $>$ EQNN $>$ QB-HNN.  The reason behind the obtained results is that NN-BP works on single solution and attempts to minimize the error iteratively while SaDE, BaDE, EQNN, and QB-HNN are optimized with an evolutionary algorithm which searches a better solution by applying  multi-dimensional exploration and exploitation throughout the population of solutions. The evolutionary optimization capabilities of BaDE with dual adaptation is superior than SaDE while LSTM provides an effective prediction accuracy but they are sensitive to random weight initialization and get overfit easily. QB-HNN and EQNN employed Qubits population along with evolutionary optimization to allow an improved intuitive learning of patterns. Further, QB-HNN is optimized with the Quantum Blackhole optimization having better exploration and exploitation capabilities with lesser hyper-parameter tuning show more accurate and precise prediction values than EQNN optimized by differential evolution algorithm. 
\begin{figure*}[!htbp]
	\centering	
	\subfigure[Web traces]{\includegraphics[width=0.31\linewidth, scale=3]{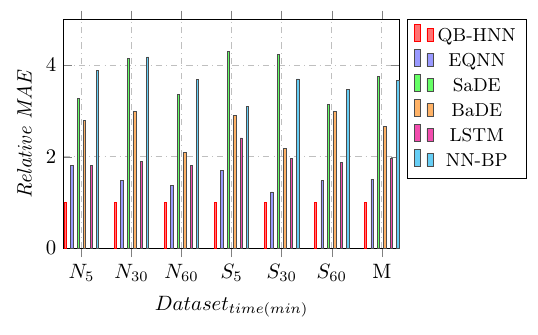}\label{ru1}} 
	\subfigure[Google Cluster traces]{\includegraphics[width=0.31\linewidth, scale=3]{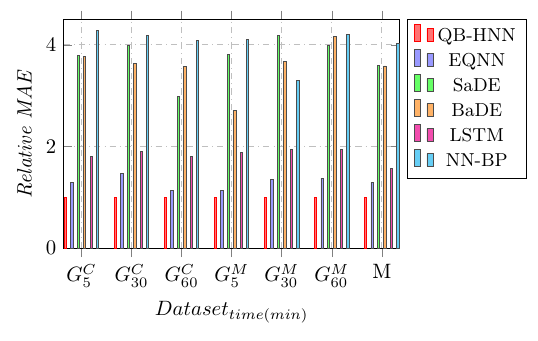}\label{ru3}} 
	\subfigure[High Performance/Grid traces]{\includegraphics[width=0.31\linewidth, scale=2]{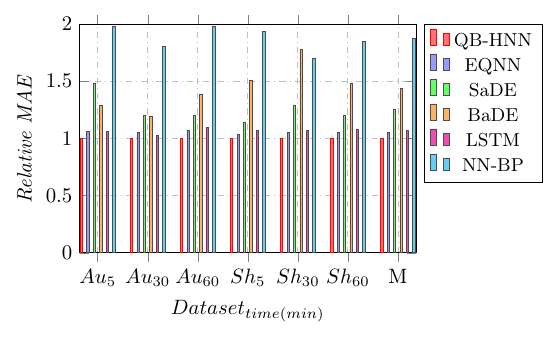}\label{GCDMin}} 	
	\caption{ {Relative MAE}}
	\label{mae}
	
\end{figure*}
\subsubsection{Relative MAE}
The performance of proposed prediction model is also compared using relative mean absolute error (MAE) (${R}el_{{MAE}}$) with different related approaches for all the datasets which is independent of scaling of data values. Eqs. (\ref{mae1}) and (\ref{mae2}) compute ${R}el_{{MAE}}$ which is the mean absolute error  of state-of-the-art method (SM) (${MAE}_{SM}$) normalized with respect to the mean absolute error  of base method i.e., QB-HNN (${MAE}_{QBHNN}$). Fig. \ref{mae} compares the efficiency of QB-HNN with relative five existing approaches for all the datasets.  The  ${MAE}$ of QB-HNN is least (i.e., 1) for all the experiments as compared with other approaches having higher ${MAE} $ values. Since the state-of-the-art methods with values of ${R}el_{{MAE}}$ closer to 1 are more acceptable, the efficiency of different comparative approaches follow the similar (as observed in MSE) trend: NN-BP $<$ SaDE $<$ BaDE $>$ LSTM $<$ EQNN $<$ QB-HNN.      

\begin{gather}
	{MAE} = {\frac{1}{m}\sum_{i=1}^{m}({D}^{Ac}-{D}^{Pr})}\label{mae1} \\
	{R}el_{{MAE}}= \frac{{MAE}_{SM}}{{MAE}_{QBHNN}} \label{mae2}
\end{gather}

{ 
\subsubsection{RMSE}

Table~\ref{tab:rmse} presents a detailed comparative evaluation of the Root Mean Square Error (RMSE), scaled by $10^{-3}$, across various benchmark techniques including NN-BP~\cite{lu2016rvlbpnn}, SaDE~\cite{kumar2018workload}, BaDE~\cite{kumar2020biphase}, CLIN~\cite{kim2020forecasting}, LSTM~\cite{kumar2018long}, EQNN~\cite{singh2021quantum}, and MCT-AQNN~\cite{gupta2024multiple}, evaluated on two representative workload traces: $G^C$ and $G^M$. The proposed QB-HNN model consistently outperforms all benchmark algorithms across prediction windows (5, 10, 30, and 60 minutes). For $G^C$ traces, QB-HNN achieves the lowest RMSE in every case, with the most significant gain at the 30-minute window (RMSE = {2.42}), outperforming the next-best (MCT-AQNN, {11.6}). For $G^M$, it yields an RMSE of {0.51} at 5 minutes and maintains superior accuracy across longer horizons. This performance stems from its quantum-inspired encoding and hybrid activation design, which efficiently capture non-linear, entangled patterns. Coupled with biologically-inspired optimization, QB-HNN achieves robust convergence and generalization, making it a reliable model for dynamic cloud workload forecasting.

\begin{table}[!htbp]
	\caption{RMSE ($ \times 10^{-3}$): QB-HNN versus state-of-the-arts } \label{tab:rmse}
	\centering
	\resizebox{0.5\textwidth}{!}{
		\begin{tabular}{cp{1.1cm}p{1.1cm}p{0.9cm}p{1.1cm}p{1.1cm}p{1.1cm}p{1.1cm}p{1.4cm}p{1.1cm}}
			\hline
			\textbf{DS} & \textbf{PWS (min)}& \textbf{{NN-BP}}\cite{lu2016rvlbpnn} &\textbf{SaDE} \cite{kumar2018workload}&\textbf{BaDE}\cite{kumar2020biphase}  & \textbf{CLIN}\cite{kim2020forecasting} &\textbf{LSTM} \cite{kumar2018long} &\textbf{EQNN}\cite{singh2021quantum} &\textbf{MCT-AQNN} \cite{gupta2024multiple}&\textbf{QB-HNN}\\ \hline
	
			\multirow{4}{*}{\rotatebox{90}{$G^C$}} &5& 18.2& 12.7& 10.4 &7.78 &19.7 &9.87& {{5.97}}& {{3.04}}\\ 
			& 10 & 21.9& 20.1& 10.9 &7.34 &13.2 &17.5& {{1.96}}& {{1.72}}\\ 
			&30	& 40.1 &35.6 & 31.1 &16.6 & 13.4&31.7& {{11.6}} & {{2.42}}\\ 
			&60	& 42.9& 38.0& 33.6 &21.7 &10.2 &20.93& {{10.1}}& {{9.92}} \\ \hline
				\multirow{4}{*}{\rotatebox{90}{$G^M$}} &5& 19.9 &11.2 &7.00& 6.85 & 19.7& 9.40 & {{0.90}} & {{0.51}} \\ 
			& 10 &37.3&	14.2	&10.7 &	7.89&	19.1 &10.5 &	{{3.73}}	&{{2.71}} \\
    			&30	&54.3	&39.4	&37.6	&15.0	&19.4	&48.5	& {{9.90}}	& {{8.89}}\\
			&60	&62.6	&48.2	&40.6	&19.6	&11.4	&39.3	 & {{10.9}}	& {{9.11}}\\ \hline	
	\end{tabular}}
\\ \footnotesize{\scriptsize DS: Dataset, $G^C$: GCD-CPU,$G^M$: GCD-Mem}
\end{table}
}
\subsubsection{Absolute Error Frequency}
\begin{figure*}[!htbp]
	\centering	
	\subfigure[Web traces]{\includegraphics[width=0.31\linewidth, scale=3]{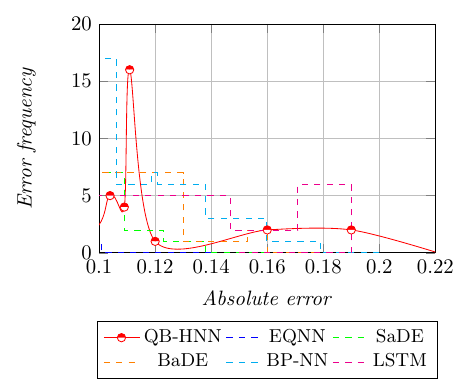}\label{ru1}} 
	\subfigure[Google Cluster traces]{\includegraphics[width=0.31\linewidth, scale=3]{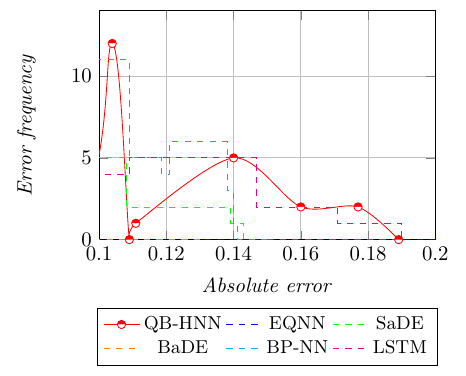}\label{ru3}} 
	\subfigure[High Performance/Grid traces]{\includegraphics[width=0.31\linewidth, scale=2]{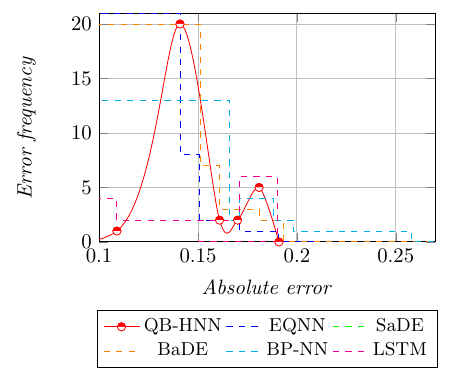}\label{GCDMin}} 	
	\caption{ {Absolute Error Frequency: Number of occurrences of absolute prediction errors within defined intervals, averaged over 30 runs.}}
	\label{aef1}
	
\end{figure*}

The prediction error achieved for QB-HNN and each of the comparative approaches is measured and analyzed by evaluating absolute prediction error and comparing its frequency for all three types of workload. Fig. \ref{aef1} compares the frequency of absolute error (\textit{Actual value} - \textit{Predicted value}), where QB-HNN produces the lowest error for most cloud workloads compared to five existing approaches. The high frequency of absolute error for the proposed QB-HNN indicates the consistent potency and the stable tendency to yield the highest prediction accuracy.

\subsubsection{ {Convergence Analysis of QB-HNN Optimization}}
\begin{figure*}[!htbp]
	\centering	
	\subfigure[Convergence]{\includegraphics[width=0.31\linewidth, scale=3]{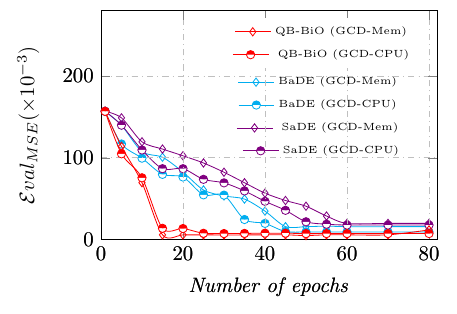}\label{ru1}} 
	\subfigure[Accuracy]{\includegraphics[width=0.31\linewidth, scale=3]{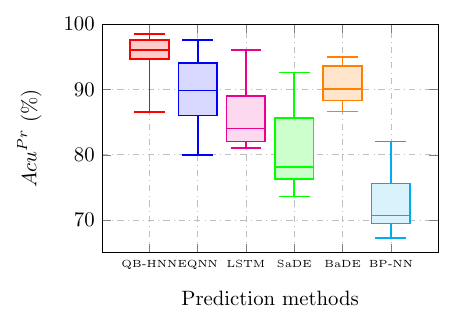}\label{ru3}} 
	\subfigure[Training Time]{\includegraphics[width=0.31\linewidth, scale=2]{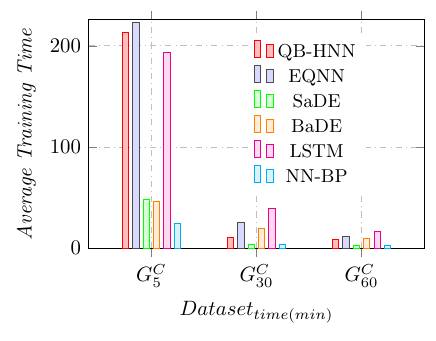}\label{GCDMin}} 	
	\caption{QB-HNN Optimization metrics}
	\label{qbhnn_optimization}
	
\end{figure*}

 {Fig. \ref{qbhnn_optimization}a compares the convergence ability of QB-BiO  algorithm (applied for optimization of QB-HNN) with convergence ability of BaDE \cite{kumar2020biphase} and SaDE \cite{kumar2018workload} for GCD-CPU ($G^C$) and GCD-Mem ($G^M$) cluster workloads. It is observed that QB-BiO  converges before 20 epochs with least values of ${E}val_{MSE}$ showing its persuasive performance over SaDE and BaDE which converges in more than 45 epochs with slightly higher values of  prediction error.} 
\par The prediction accuracy of proposed and existing state-of-the-art approaches is compared  with the help of box-plots in Fig. \ref{qbhnn_optimization}b. A boxplot distributes $Acu^{Pr} (\%)$ statistically via quartiles where bottom, middle and top of the box are the first, second, and third quartiles specifying the average of median range of values achieved for the  prediction accuracy by applying the respective approach for $G^C$. The ends of the whiskers are the least and highest of all values of prediction accuracy, respectively.
\par The average training-time of the QB-HNN is compared with the five comparative approaches is shown in Fig. \ref{qbhnn_optimization}c for $G^C$ cluster workload. The highest time is observed for EQNN due to the generation of qubit-based network weights and least for NN-BP which shows least prediction accuracy among all the approaches. However, the durability and applicability of the proposed predictor is independent of the impact of the training time because the training is a periodic task and can be executed concurrently on the servers equipped with sufficient computing resources.

{  
\subsubsection{Sensitivity Analysis} 

Table~\ref{table:Sensitivity Analysis} presents the sensitivity analysis of the proposed QB-BiO algorithm under varying population sizes ($Z$) and qubit cluster counts ($K$).  MSE values are evaluated for both training and testing phases to assess the generalization ability and stability of the algorithm under different parameter settings. Notably, the configuration with $K = 5$ and $Z = 25$ achieves the lowest testing MSE of $0.00412$, suggesting an optimal balance between search space exploration and exploitation. This indicates that a moderate population size enhances the convergence quality without overfitting or increasing computational cost. In contrast, smaller population sizes such as $Z = 15$ (with $K = 5$) yield higher testing MSE ($0.00551$), implying insufficient diversity and premature convergence. Similarly, a large cluster count (e.g., $K = 6$ with $Z = 36$) leads to a higher training MSE ($0.06404$), possibly due to over-clustering that introduces noisy decision boundaries in the solution space. These results empirically validate that the performance of QB-BiO is highly sensitive to the choice of qubit clusters and population size, and a careful balance is required to optimize its learning and generalization capacity.

\begin{table}[!htbp]
	\centering
	\caption[Table caption text] {Sensitivity Analysis across varying population size}  
	\label{table:Sensitivity Analysis}
		\resizebox{6cm}{!}{
	\begin{tabular}{|p{1.5cm}|p{1.2cm}|p{1.2cm}| p{1.2cm}|}
		\hline			
		\textbf{Qubit Clusters\# (K)}&  	\textbf{Population size (Z)} &\textbf{MSE (training)} & \textbf{MSE (testing)} \\ 	\hline
			
		4 & 16 & 0.05404 & 0.00414 \\ 	\hline
5 & 25 & 0.01804 & 0.00412 \\ 	\hline

5 & 15 & 0.01903 & 0.00551 \\ 	\hline
5 & 20 & 0.02237 & 0.00441 \\ 	\hline
6 & 36 & 0.06404 & 0.00467 \\ 	\hline      
	\end{tabular}}
\end{table}

}

\subsubsection{Computational Complexity}
The comparison of  computational complexity of learning process of proposed and existing prediction models is depicted in Table \ref{table:PLcomparison}, wherein ${N}$ represents number of input neurons, ${Z}$ is the size of initial population of networks, ${K}$ is maximum number of epochs, ${M}$ is number of training data samples, ${N}_c$ and ${N}_o$ are number of memory cells and number of output units, and $t$ is total number of prediction time-interval. It is reported that QB-HNN has computational complexity equivalent to SaDE \cite{kumar2018workload} and BaDE\cite{kumar2020biphase} which is greater than the complexity of NN-BP. The comparison furnishes that the proposed QB-HNN is durable and potent prediction model with higher accuracy with acceptable computational complexity of the intended learning algorithm i.e., QB-BiO .   
\begin{table}[!htbp]
	\centering
	\caption[Table caption text] {Comparison of Computational Complexity}  
	\label{table:PLcomparison}
		\resizebox{5.7cm}{!}{
	\begin{tabular}{|l| c|}
		\hline			
		\textbf{Approach}&  	\textbf{Computational Complexity} \\ 	\hline
		QB-HNN (proposed)  & ${O}({Z\times K}\times {N}^2 \times {M})$\\ \hline	
		EQNN \cite{singh2021quantum} &  ${O}({N}^2\times {D} \times {Z} \times {K})$\\ \hline
		SaDE \cite{kumar2018workload} & ${O}({K} \times {M} \times {N}^2 \times {Z})$ \\ \hline
		BaDE \cite{kumar2020biphase} &${O}({K} \times {M} \times {N}^2 \times {Z})$  \\ \hline
		LSTM \cite{kumar2018long}& $ {O}({N}_c \times {N}_c + {N}_c\times {N}_o)$ \\ \hline
		NN-BP \cite{prevost2011prediction}& ${O}({K} \times {M} \times {N}^2)$\\ \hline	
			
	\end{tabular}}
\end{table}

{ \subsection{Adaptability, Scalability, and Deployment Feasibility}

The proposed QB-HNN learning framework is inherently designed for adaptability and scalability, making it well-suited for deployment in large-scale, heterogeneous cloud datacenter environments. A key strength of the framework lies in its ability to cope with dynamic load variations driven by external factors such as hardware failures, unpredictable user demands, and network-level disruptions. This adaptability is facilitated by the biphase nature of QB-BiO, which balances exploration and exploitation, enabling dynamic re-optimization of the qubit-based synaptic weights and bias parameters in response to fluctuating resource states. Furthermore, the QB-HNN model supports incremental retraining using real-time workload traces and incorporates error-tolerant feedback loops to enhance resilience against transient failures, ensuring continuity and robustness in volatile cloud ecosystems.

In terms of scalability and industrial deployment feasibility, although the model demonstrates promising convergence behavior and learning efficiency under simulated conditions, real-world integration poses several practical challenges. One key concern is the seamless interoperability of quantum-inspired optimization techniques with conventional cloud management tools such as Kubernetes, OpenStack, and proprietary orchestration platforms. Bridging this gap may require the development of middleware abstractions or custom plugin interfaces to facilitate integration without overhauling existing infrastructure. Additionally, the representation of dynamic cloud workloads through qubit-based encodings can impose computational and memory overhead on classical systems lacking native quantum acceleration. To mitigate these limitations, the framework advocates for modular and distributed deployment strategies, hierarchical optimization workflows, and parallel execution of QB-BiO across multi-tenant cloud environments. These design choices aim to ensure robust scalability, efficient resource utilization, and minimal disruption to existing cloud operations, thus laying a practical foundation for industrial adoption.

}

\section{Conclusions and Future Work}
A novel QB-HNN prediction model is proposed to forecast an extensive range of dynamic cloud workloads with precise accuracy. This model intelligently integrates quantum computing mechanics with machine learning, enhancing the learning capacity of the neural network by utilizing a diverse population of qubits as neural weight connections. The Hadamard gate serves as an activation function, allowing superposition of neural qubits at hidden and output layers. Optimized with a newly developed QB-BiO optimization algorithm, the model facilitates persuasive learning of extracted patterns. The precision of the proposed prediction model was extensively evaluated using six real-world benchmark datasets in three categories and compared with five state-of-the-art methods. The results confirm the superior performance of QB-HNN, demonstrating improved prediction accuracy over existing approaches. This work opens new avenues for leveraging the computational properties of quantum mechanics in conjunction with machine learning models.  {Future research will focus on enhancing the Hadamard Quantum Neural Network by integrating transfer learning and multi-step task learning mechanisms to improve its accuracy, robustness, and adaptability to unseen and evolving cloud workloads.} Additionally, efforts will be directed toward strengthening the framework’s interoperability with existing DevOps pipelines and exploring hybrid quantum-classical simulation environments, thereby facilitating seamless integration and deployment in real-world industrial cloud infrastructures.

\bibliographystyle{IEEEtran}

\bibliography{bibfile}

\begin{IEEEbiography}[{\includegraphics[width=0.9\linewidth]{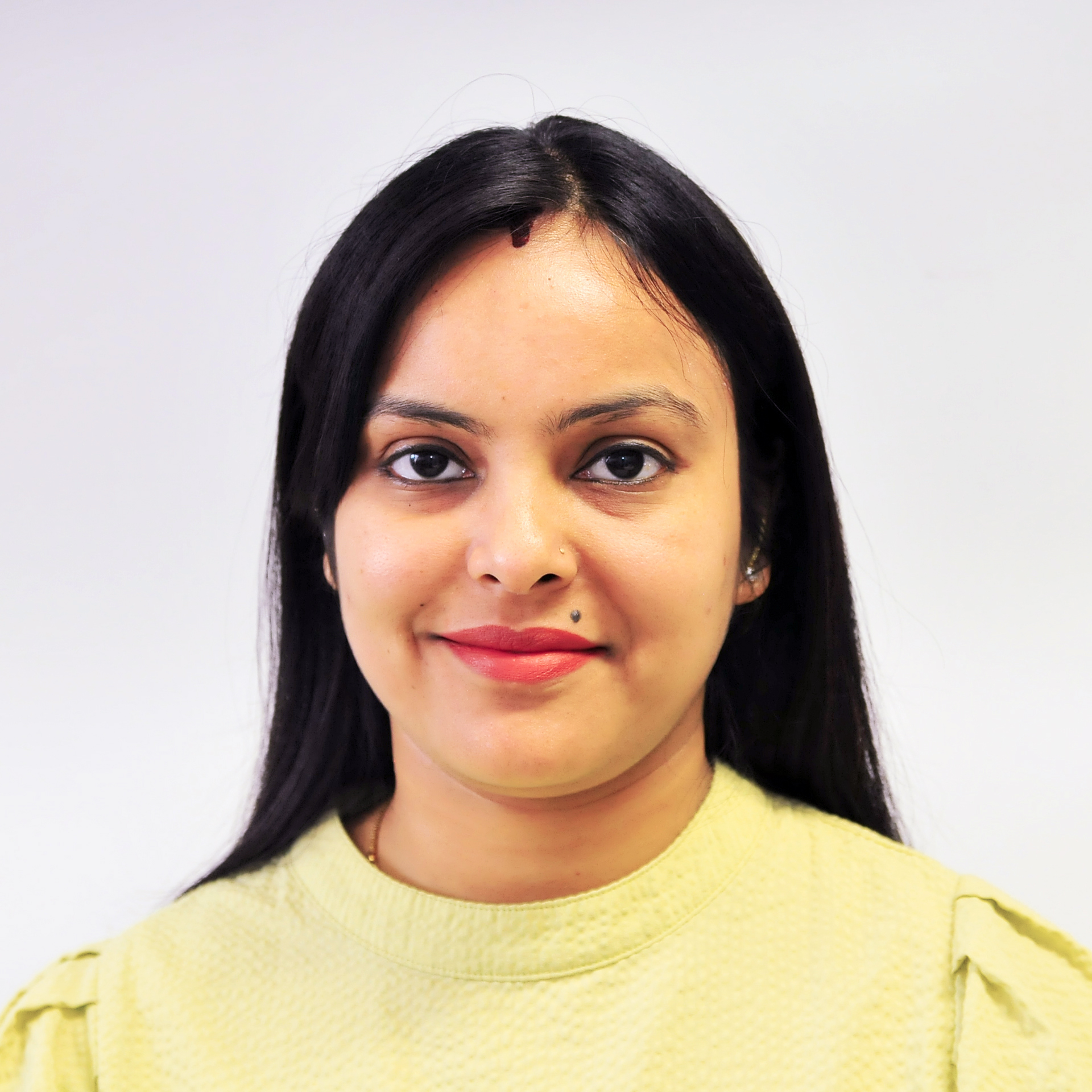}}]{Deepika Saxena}
is working as an Associate Professor in the Division of Information Systems at the University of Aizu, Japan.  She received her Ph.D. degree in Computer Science from the National Institute of Technology, Kurukshetra, India, and completed her Post Doctorate from the Department of Computer Science at Goethe University, Frankfurt, Germany.  She is the recipient of the prestigious IEEE TCSC Early Career Researcher Award 2024, IEEE TCSC 2023 Outstanding Ph.D. Dissertation Award and EUROSIM 2023 Best Ph.D. Thesis Award. She is the recipient of the prestigious JSPS KAKENHI Early Career Young
Scientist Research Grant FY2024.  Also, her research paper, published in the IEEE Transactions on Cloud Computing Journal, received the 2022 Best Paper Award. Her major research interests include Neural networks, Evolutionary algorithms, Resource management and Security in Cloud Computing, Internet traffic management, and Quantum machine learning, Dynamic Caching Management.
\end{IEEEbiography}
\vskip 0pt plus -1fil 
\begin{IEEEbiography}[{\includegraphics[width=0.7\linewidth]{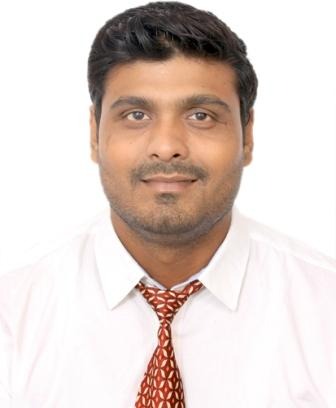}}]{Hari Mohan Gaur} earned his
Ph.D from National Institute of Technology Kurukshetra  in Reversible and Quantum Computation. He has more than 15 years of experience in academic, research and administrative capacities. Currently, he is working as a Postdoctoral researcher in Institute of Computer Science at Goethe University, Germany.  He is a distinguished Researcher, well known in Academic Fraternity for his interdisciplinary research in the areas of Quantum Computation, Fault Tolerant Digital Design, Data Security in Cloud Environment.  	
\end{IEEEbiography}
\vskip 0pt plus -1fil 
\begin{IEEEbiography}[{\includegraphics[width=0.7\linewidth]{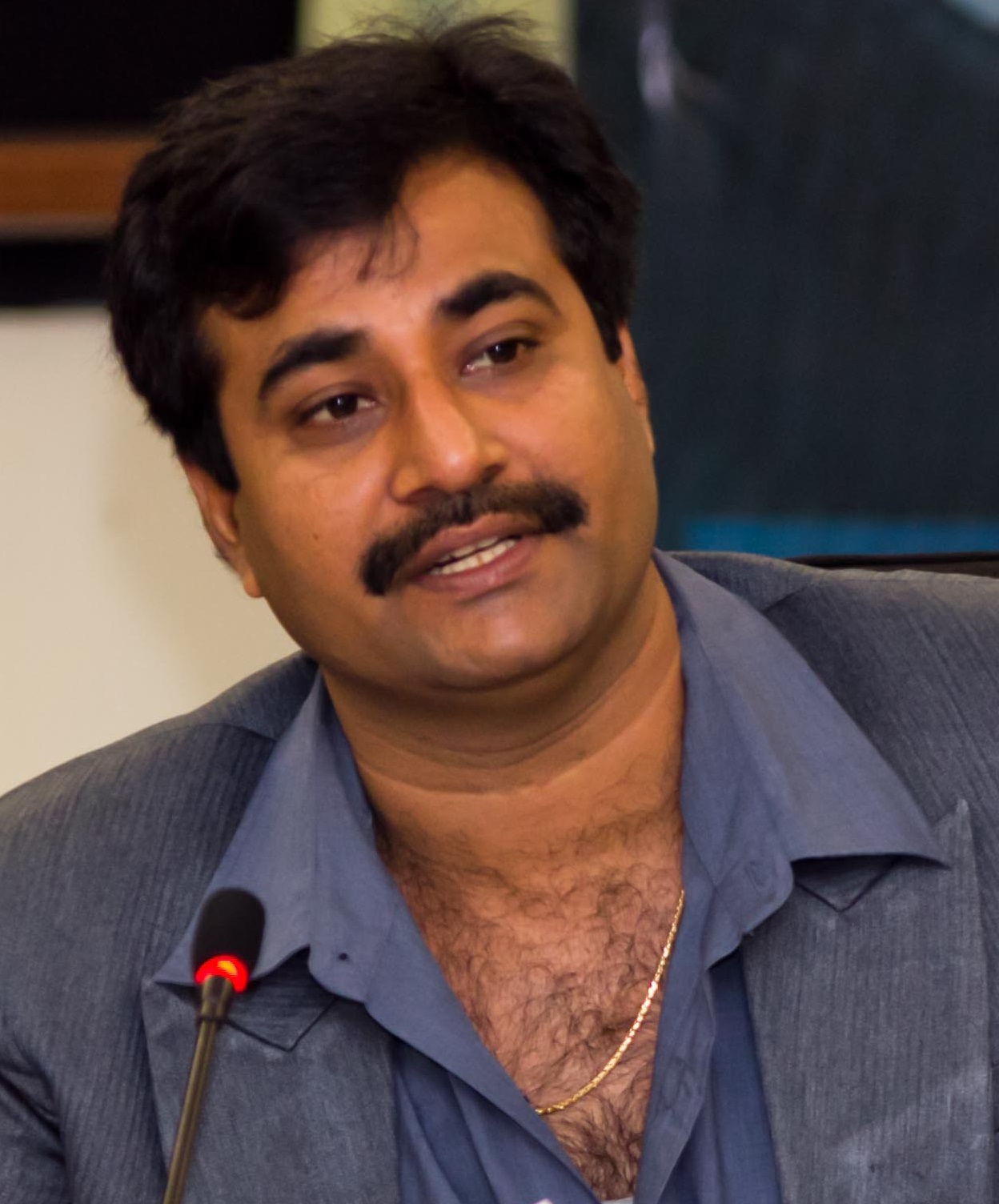}}]{Ashutosh Kumar Singh} (Senior Member, IEEE) is working as a Professor and Director of Indian Institute of Information Technology Bhopal, India. Also, he is working as Adjunct Professor in the University of Economics and Human Sciences, Warsaw, Poland. He received his Ph.D. in Electronics Engineering from Indian Institute of Technology, BHU, India and Post Doc from Department of Computer Science, University of Bristol, UK. He has research and teaching experience in various Universities of the India, UK, and Malaysia. His research area includes Design and Testing of Digital Circuits, Data Science, Cloud Computing, Machine Learning, Security. He has published more than 400 research papers in different journals and conferences of high repute.   His research paper, published in the IEEE Transactions on Cloud Computing Journal, was honored with the 2022 Best Paper Award by the IEEE Computer Society Publications Board. 

\end{IEEEbiography}
\vskip 0pt plus -1fil 
\begin{IEEEbiography}[{\includegraphics[width=0.7\linewidth]{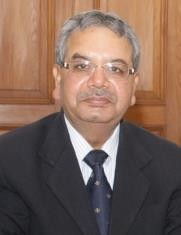}}]{Anand Mohan}
	is a Professor of Electronics Engineering at Institute of Technology, Banaras Hindu University where he has held as several important administrative positions namely Member of Executive Council, Head of the Department of Electronics Engineering, Coordinator, Centre for Research in Microprocessor Applications (established by MHRD), and In charge, University Science Instrumentation Centre. He has 35 years rich experience of serving both academia and industry in various capacities. He obtained Ph.D., PG, and UG degrees in Electronics Engineering from Banaras Hindu University in 1994, 1977, and 1973 respectively. He has made notable contributions to the academic and research development in Electronics Engineering at Banaras Hindu University by creating dedicated research groups of eminent academic experts from the country and abroad. He conducted high quality research in the emerging areas like fault tolerant/survivable system design, information security, and embedded systems.
\end{IEEEbiography}

\end{document}